\documentclass[12pt]{article}
\usepackage{amssymb,amsmath,amsbsy}
\usepackage{dsfont}
\usepackage{cite}
\usepackage[title]{appendix}
\usepackage{mathrsfs}
\usepackage[unicode=true,pdfencoding=auto]{hyperref}
\usepackage[all]{hypcap}       
\makeatletter
\g@addto@macro\bfseries{\boldmath}
\makeatother	

\DeclareMathOperator{\Tr}{Tr}

\DeclareTextCommand{\uppercaseLambda}{PU}{\83\233}
\DeclareTextCommand{\textzeta}{PU}{\83\266}
\DeclareTextCommand{\texttheta}{PU}{\83\270}
\DeclareTextCommand{\textsigma}{PU}{\83\303}
\DeclareTextCommand{\textmu}{PU}{\83\274}
\DeclareTextCommand{\textnu}{PU}{\83\275}
\DeclareTextCommand{\overbar}{PU}{\83\005}
\DeclareTextCommand{\overarrow}{PU}{\9040\327}
\DeclareTextCommand{\superu}{PU}{\9035\130}
\DeclareTextCommand{\subv}{PU}{\9035\145}
\DeclareTextCommand{\uppercasepsi}{PU}{\83\250}
\DeclareTextCommand{\texteta}{PU}{\83\267}
\DeclareTextCommand{\textchi}{PU}{\83\307}
\DeclareTextCommand{\textgamma}{PU}{\83\263}

\newcommand{\mathbold}[1]{\mbox{\boldmath $#1$}}
\def\ifmath#1{\relax\ifmmode #1\else $#1$\fi}
\def\half{\tfrac{1}{2}}
\def\quarter{\tfrac{1}{4}}
\def\ls#1{\ifmath{_{\lower1.5pt\hbox{$\scriptstyle #1$}}}}
\def\lsup#1{\ifmath{^{\lower3pt\hbox{$\scriptstyle #1$}}}}
\def\phm{\phantom{-}}
\def\eps{\epsilon}

\def\iso{\mathchoice{\cong}{\cong}{\isoS}{\cong}}
\def\isoS{\vbox{\baselineskip 0pt  \lineskip 0.5pt
    \ialign{$ \mathsurround=0pt  \scriptstyle \hfil ## \hfil $\crcr
        \sim \crcr = \crcr}}}

\catcode`~=11 
\newcommand{\urltilde}{\kern -.15em\lower .7ex\hbox{~}\kern .04em}
\catcode`~=13 

\newenvironment{Eqnarray}%
     {\arraycolsep 0.14em\begin{eqnarray}}{\end{eqnarray}}
\newcommand{\beqa}{\begin{Eqnarray}}
\newcommand{\eeqa}{\end{Eqnarray}}
\def\phm{\phantom{-}}
\def\T{{\mathsf T}}

\def\axis{\boldsymbol{\hat n}}
\def\beq{\begin{equation}}
\def\eeq{\end{equation}}
\def\eq#1{eq.~(\ref{#1})}
\def\Eq#1{Eq.~(\ref{#1})}
\def\Eqs#1#2{Eqs.~(\ref{#1}) and (\ref{#2})}
\def\eqs#1#2{eqs.~(\ref{#1}) and (\ref{#2})}
\def\eqss#1#2#3{eqs.~(\ref{#1}), (\ref{#2}) and (\ref{#3})}
\def\Eqss#1#2#3{Eqs.~(\ref{#1}), (\ref{#2}) and (\ref{#3})}
\def\eqst#1#2{eqs.~(\ref{#1})--(\ref{#2})}

\def\half{\tfrac{1}{2}}
\def\newcdot{\kern.06em{\cdot}\kern.06em}
\def\sigmabar{\overline\sigma}
\def\id{\boldsymbol{I}}
\def\lsub#1{_{\lower 1.5pt\hbox{$\scriptstyle#1$}}}

\def\T{{\mathsf T}}
\def\pmat#1{\begin{pmatrix}#1\end{pmatrix}}
\renewcommand{\thefootnote}{\fnsymbol{footnote}}

\begin{document}
\begin{flushright}
\normalsize{
December, 2023
}
\end{flushright}
\vspace{1cm}

\begin{center}
\Large\bf\boldmath
Explicit form for the most general Lorentz transformation revisited

\unboldmath
\end{center}
\vspace{0.4cm}
\begin{center}

Howard E.~Haber \!\footnote{Electronic address: haber@scipp.ucsc.edu}\\
 {\sl Santa Cruz Institute for Particle Physics \\
 University of California, 1156 High Street, Santa Cruz, CA 95064 USA}\\[0.2cm]
\end{center}

\vspace{0.2cm}

\renewcommand{\thefootnote}{\arabic{footnote}}
\setcounter{footnote}{0}

\begin{abstract}
Explicit formulae for the $4\times 4$ Lorentz transformation matrices corresponding to a pure boost and a pure three-dimensional rotation are very well-known.   Significantly less well-known is the explicit formula for a general Lorentz transformation with arbitrary nonzero boost and rotation parameters.  We revisit this more general formula by presenting two different derivations.  The first derivation (which is somewhat simpler than previous ones appearing in the literature) evaluates the exponential of a $4\times 4$ real matrix $A$, where $A$ is a product of the diagonal matrix ${\rm diag}(+1, -1, -1, -1)$ and an arbitrary $4\times 4$ real antisymmetric matrix.  The formula for $\exp A$ depends only on the eigenvalues of $A$ and makes use of the Lagrange interpolating polynomial.   The second derivation exploits the observation that the spinor product $\eta^\dagger\sigmabar^\mu\chi$ transforms as a Lorentz four-vector, where $\chi$ and~$\eta$ are two-component spinors.  The advantage of the latter derivation is that the corresponding formula for a general Lorentz transformation $\Lambda$ reduces to the computation of the trace of a product of $2\times 2$ matrices.  Both computations are shown to yield equivalent expressions for $\Lambda$.

\end{abstract}
\clearpage

\section{Introduction}
\label{realintro}

In the theory of special relativity, space and time are combined into Minkowski spacetime (e.g., see Ref.~\cite{Sexl}).   
Two different inertial reference frames (with coinciding origins fixed) are related through a Lorentz transformation.  Equivalently, consider
a four-vector, $\boldsymbol{x}=(x^0\,;\,\boldsymbol{\vec{x}})$, with squared-length $\|\boldsymbol{x}\|^2\equiv\eta_{\mu\nu}x^\mu x^\nu=(x^0)^2-|\boldsymbol{\vec{x}}|^2$ (with an implied double sum over the repeated indices $\mu,\nu\in\{0,1,2,3\}$), where $\eta_{\mu\nu}={\rm diag}(+1,-1,-1,-1)$ is the Minkowski spacetime metric.   One can also define the Lorentz 
transformation $\Lambda$ as a symmetry transformation of a four-vector, $\boldsymbol{x'}=\Lambda\boldsymbol{x}$, that preserves the length of $\boldsymbol{x}$.   Since the length of a four-vector is a scalar quantity and thus invariant under a Lorentz transformation, it follows that 
$\eta_{\alpha\beta}=\Lambda^\mu{}_\alpha\Lambda^\nu{}_\beta\eta_{\mu\nu}$, which serves as the general definition of the $4\times 4$ Lorentz transformation matrix [cf.~\eqst{etavv}{lambdarelation}].  Moreover, this same equation implies that $\eta_{\mu\nu}$ is an invariant tensor.
Indeed, the Lorentz transformations (along with spacetime translations) are the maximally allowed symmetry transformations of Minkowski spacetime in which the spacetime metric is left invariant (e.g., see Ref.~\cite{Markoutsakis}).

Consider two inertial reference frames with coinciding origins, where one reference frame is moving with respect to the other with three-vector velocity~$\boldsymbol{\vec{v}}$.
The corresponding Lorentz transformation is called a Lorentz boost.  The
boost parameters are defined by the components of the three-vector $\boldsymbol{\vec{\zeta}}\equiv(\boldsymbol{\vec{v}}/v)\tanh^{-1}(v/c)$, where $v\equiv |\boldsymbol{\vec{v}}|$ and $c$ is the speed of light.  
However, this is not the most general Lorentz transformation.  For example, let $R$ be an arbitrary $3\times 3$ orthogonal matrix of unit determinant, i.e., a proper rotation matrix parametrized by 
the components of the three-vector $\boldsymbol{\vec{\theta}}\equiv \theta\boldsymbol{\hat{n}}$ (such that $\theta$ is the angle of rotation, counterclockwise, about a fixed axis
that lies along the unit vector $\boldsymbol{\hat{n}}$).  Then, 
the transformation $x^{\prime\,0}=x^0$ and $\boldsymbol{\vec{x}^{\,\prime}}=R\boldsymbol{\vec{x}}$  is also a Lorentz transformation as it leaves the Minkowski spacetime metric invariant.
The corresponding matrix representations of the general Lorentz boost and three-dimensional rotation are quite well known [see \eqs{Lmatrixzeta}{Lmatrix2}, respectively] and are reviewed in Section~\ref{intro}.

A more general Lorentz transformation matrix, which shall henceforth be denoted by  
$\Lambda(\boldsymbol{\vec{\zeta}}\,,\,\boldsymbol{\vec{\theta}})$,  corresponds to a simultaneous boost and rotation.
As shown in Section~\ref{sec:L}, $\Lambda(\boldsymbol{\vec{\zeta}}\,,\,\boldsymbol{\vec{\theta}})$ can be expressed as 
 the exponential of a $4\times 4$ matrix,
\beq \label{fourxfour}
\Lambda(\boldsymbol{\vec{\zeta}}\,,\,\boldsymbol{\vec{\theta}})=\exp \begin{pmatrix} 0 &\phm \zeta^1 &\phm \zeta^2 &\phm \zeta^3 \\
\zeta^1 & \phm\! 0 & -\theta^3 &\phm \theta^2 \\\zeta^2 & \phm\theta^3 & \phm\! 0 & -\theta^1 \\
\zeta^3 & -\theta^2 &\phm \theta^1 & \phm\! 0\end{pmatrix}.
\eeq
In contrast to $\Lambda(\boldsymbol{\vec{\zeta}}\,,\,\boldsymbol{\vec{0}})$ and $\Lambda(\boldsymbol{\vec{0}}\,,\,\boldsymbol{\vec{\theta}})$, which correspond to 
a Lorentz boost matrix and a three-dimensional rotation, respectively, an explicit form for $\Lambda(\boldsymbol{\vec{\zeta}}\,,\,\boldsymbol{\vec{\theta}})$
is much less well known.  

The first
published formula for $\Lambda(\boldsymbol{\vec{\zeta}}\,,\,\boldsymbol{\vec{\theta}})$ appeared in Ref.~\cite{ZR}.  Subsequent derivations have also been given in Refs.~\cite{Geyer,DM,AR}.
These derivations are based on the Cayley-Hamilton theorem of linear algebra (e.g., see Section 8.4 of Ref.~\cite{Carrell}), which asserts that any $n\times n$ matrix $A$ satisfies its own characteristic equation, $p(x)=\det(A-x\boldsymbol{I}_n)=0$, where $\boldsymbol{I}_n$ is the $n\times n$ identity matrix and $p(x)$ is an $n$th-order polynomial whose roots are the eigenvalues of $A$.  
That is, $p(A)$ is equal to the zero matrix.  It
follows that for any integer $k\geq n$, the matrix $A^k$ can be expressed as a linear combination of $\boldsymbol{I}_n$, $A$, $A^2, \ldots A^{k-1}$.  In particular, 
\beq \label{cs}
\Lambda(\boldsymbol{\vec{\zeta}}\,,\,\boldsymbol{\vec{\theta}})\equiv \exp A=\sum_{k=0}^\infty \frac{A^k}{k!}=c_0\boldsymbol{I}_4+c_1 A+ c_2 A^2+c_3 A^3\,,
\eeq
where each of the coefficients $c_k$ is an infinite series whose terms depend on the eigenvalues of $A$.  
Note that by setting either $\boldsymbol{\vec{\theta}}=\boldsymbol{\vec{0}}$ or $\boldsymbol{\vec{\zeta}}=\boldsymbol{\vec{0}}$ 
 in \eq{fourxfour}, one can easily compute the resulting matrix exponential to derive the well-known expressions given in \eqs{Lmatrixzeta}{Lmatrix2}, respectively.  In contrast, 
 if both the boost vector and the rotation vector are nonzero, then the corresponding computation of the matrix exponential, which is carried out in Refs.~\cite{ZR,Geyer}, is significantly more difficult.   
In Ref.~\cite{DM}, this computation is performed by showing that a Lorentz transformation matrix $g$ exists such that the $4\times 4$ matrix $\widetilde{A}\equiv gAg^{-1}$ in block matrix form is made up of very simple $2\times 2$ matrix blocks.  The exponential $\exp\widetilde{A}$ is then easy to evaluate directly via its Taylor series to obtain the coefficients $c_k$, and
\beq
\exp A=g^{-1}(\exp\widetilde{A})g=g^{-1}\bigl[c_0\boldsymbol{I}_4+c_1 \widetilde{A}+ c_2 \widetilde{A}^{\,2}+c_3 \widetilde{A}^{\,3}\bigr]g=c_0\boldsymbol{I}_4+c_1 A+ c_2 A^2+c_3 A^3\,.
\eeq
Finally, Ref.~\cite{AR} derives a system of four linear equations for the coefficients $c_k$ in \eq{cs}, whose solution provides the desired expression for $\exp A$.
Similar techniques have also been employed in Ref.~\cite{Kaiser} to obtain an explicit expression for the matrix exponential function of SO(4).

In this paper, we shall provide a somewhat simpler and more straightforward evaluation of $\Lambda(\boldsymbol{\vec{\zeta}}\,,\,\boldsymbol{\vec{\theta}})$ as compared to the derivations given in Refs.~\cite{ZR,Geyer,DM,AR}.    
In Section~\ref{intro}, we first exhibit the explicit forms for the general Lorentz boost and the three-dimensional rotation matrices of Minkowski spacetime, which correspond to special cases of the more general $4\times 4$ Lorentz transformation matrix, as noted above.
In Section~\ref{sec:L}, an expression for the most general Lorentz transformation is then derived.  Indeed, it is sufficient to consider the set of all Lorentz transformations that are continuously connected to the identity, known as the proper orthochronous Lorentz transformations (e.g., see Ref.~\cite{Sexl}).   The matrix representation of any element of this latter set can be expressed in the form given by \eq{fourxfour}, as discussed below \eq{exponentiate}.
In Section~\ref{sec:fourbyfour}, we explicitly evaluate  \eq{fourxfour} for arbitrary boost and rotation parameters.  
We then demonstrate that an alternative derivation of $\Lambda(\boldsymbol{\vec{\zeta}}\,,\,\boldsymbol{\vec{\theta}})$ can be given that only
involves the manipulation of $2\times 2$ matrices, by making use of two-component spinors.  
In particular, we show in Section~\ref{sec:twobytwo} that the most general proper orthochronous Lorentz transformation matrix can be expressed as a trace of the product of four $2\times 2$ matrices, which is then explicitly evaluated.  Both methods for computing $\Lambda(\boldsymbol{\vec{\zeta}}\,,\,\boldsymbol{\vec{\theta}})$ are carried out in pedagogical detail.  In Section~\ref{sec:reconcile}, we check that both computations yield the same expression for $\Lambda(\boldsymbol{\vec{\zeta}}\,,\,\boldsymbol{\vec{\theta}})$.   
Final remarks are presented in Section~\ref{final}, and some related discussions are relegated to the appendices.

\section{Lorentz transformations--special cases}
\label{intro}

In a first encounter with special relativity, a student learns how the spacetime coordinates change between two inertial reference frames $K$ and $K'$.  If the spacetime coordinates with respect to $K$ are $(ct;x,y,z)$ and
the spacetime coordinates with respect to $K'$ are $(ct';x',y',z')$, where $K'$ is moving relative to $K$ with velocity $\boldsymbol{\vec{v}}=v\boldsymbol{\hat{x}}$ in the $x$ direction, then 
\beqa
ct'&=&\gamma(ct-\beta x)\,,\label{e1}\\
x'&=&\gamma(x-\beta ct)\,,\label{e2}\\
y'&=&y\,,\label{e3}\\
z'&=&z\,,\label{e4}
\eeqa
where $c$ is the speed of light and
\beq
\beta\equiv\frac{v}{c}\,,\qquad\quad \gamma\equiv(1-\beta^2)^{-1/2}\,.
\eeq

It is straightforward to generalize the above results for an arbitrary velocity $\boldsymbol{\vec{v}}$ by writing
\beq
\boldsymbol{\vec{x}}=\boldsymbol{\vec{x}_\parallel}+\boldsymbol{\vec{x}\ls\perp}\,,
\eeq 
where $\boldsymbol{\vec{x}_\parallel}$ is the projection of $\boldsymbol{\vec{x}}$ along the direction of $\boldsymbol{\vec{v}}\equiv c\boldsymbol{\vec{\beta}}$, and 
$\boldsymbol{\vec{x}\ls\perp}$ is perpendicular to~$\boldsymbol{\vec{v}}$ (so that $\boldsymbol{\vec{x}_\parallel\newcdot\vec{x}\ls\perp}=0$).  The definition of $\boldsymbol{\vec{x}_\parallel}$ implies that
\beq \label{xb}
\frac{\boldsymbol{\vec{x}_\parallel}}{|\boldsymbol{\vec{x}_\parallel}|}=\frac{\boldsymbol{\vec{\beta}}}{\beta}\,,
\eeq
where $\beta\equiv|\boldsymbol{\vec{\beta}}|$.  Note that $0\leq\beta<1$ for any particle of non-zero mass.

In light of \eq{xb}, \eqst{e1}{e4} are equivalent to
\beqa
ct'&=&\gamma(ct-\boldsymbol{\vec{\beta}\newcdot\vec{x}_\parallel})\,,\label{e5}\\
\boldsymbol{\vec{x}_\parallel}^{\!\!\!\prime}&=&\gamma(\boldsymbol{\vec{x}_\parallel}-\boldsymbol{\vec{\beta}}ct)\,,\label{e6}\\
\boldsymbol{\vec{x}_\perp}^{\!\!\!\!\prime}&=& \boldsymbol{\vec{x}_\perp}\,,\label{e7}
\eeqa
where $\gamma\equiv (1-|\boldsymbol{\vec{\beta}}|^2)^{-1/2}$.  Note that $1\leq\gamma<\infty$ for any particle of non-zero mass.
More explicitly,
\beq \label{xpp}
\boldsymbol{\vec{x_\parallel}}=\left(\frac{\boldsymbol{\vec{\beta}\newcdot\vec{x}}}{\beta^2}\right)\boldsymbol{\vec{\beta}}\,,\qquad\quad
\boldsymbol{\vec{x}\ls\perp}=\boldsymbol{\vec{x}}-\left(\frac{\boldsymbol{\vec{\beta}\newcdot\vec{x}}}{\beta^2}\right)\boldsymbol{\vec{\beta}}\,,
\eeq
which yield $\boldsymbol{\vec{\beta}\newcdot\vec{x}_{\parallel}}=\boldsymbol{\vec{\beta}\newcdot \vec{x}}$ and $\boldsymbol{\vec{\beta}\newcdot\vec{x}\ls\perp}=0$ as required.
Inserting the expressions given in \eq{xpp} back into \eqst{e5}{e7}, we end up with the well-known result (e.g., see eq.~(11.19) of Ref.~\cite{Jackson}):
\beqa
ct'&=& \gamma(ct-\boldsymbol{\vec{\beta}\newcdot\vec{x}})\,,\label{e8}\\
\boldsymbol{\vec{x}}^{\,\prime}&=& \boldsymbol{\vec{x}}+\frac{(\gamma-1)}{\beta^2}(\boldsymbol{\vec{\beta}\newcdot\vec{x}})\boldsymbol{\vec{\beta}}-\gamma \boldsymbol{\vec{\beta}}ct\,.\label{e9}
\eeqa

Following eq.~(11.20) of Ref.~\cite{Jackson}, it is convenient to introduce the boost parameter $\zeta$ (also called the rapidity),
\beq \label{betagamma}
\gamma=\cosh\zeta\,,\qquad\quad \gamma\beta=\sinh\zeta\,,
\eeq
since the definitions of $\beta$ and $\gamma$ are consistent with the relation $\cosh^2\zeta-\sinh^2\zeta=1$.  In particular, note that $0\leq\zeta<\infty$.
We then define the \textit{boost vector} $\boldsymbol{\vec{\zeta}}$ to be the vector of magnitude $\zeta$ that points in the direction of $\boldsymbol{\vec{\beta}}$.  Since \eq{betagamma} yields $\beta=\tanh\zeta$, it follows that 
\beq \label{boostv}
\boldsymbol{\vec{\zeta}}\equiv \frac{\boldsymbol{\vec{\beta}}}{\beta}\tanh^{-1}\beta\,.
\eeq
In terms of the boost vector $\boldsymbol{\vec{\zeta}}$ and its magnitude $\zeta\equiv |\boldsymbol{\vec{\zeta}}|$, \eqs{e8}{e9} yield
\beqa
ct'&=& ct\,\cosh\zeta
-\frac{\boldsymbol{\vec{\zeta}\newcdot\vec{x}}}{\zeta}\sinh\zeta\,, \label{e10} \\
\boldsymbol{\vec{x}}^{\,\prime}&=&\boldsymbol{\vec{x}}- \frac{\boldsymbol{\vec{\zeta}}}{\zeta}\left[ct\,\sinh\zeta-\frac{\boldsymbol{\vec{\zeta}\newcdot\vec{x}}}{\zeta}
(\cosh\zeta-1)\right]\,.\label{e11}
\eeqa

Before proceeding, it is instructive to distinguish between active and passive Lorentz transformations (e.g., see Ref.~\cite{Sexl}).   The Lorentz transformation discussed above is a \textit{passive} transformation, since the reference frame $K$ (specified by the coordinate axes) is transformed into~$K'$, while leaving the observer fixed.   Equivalently, one can consider an \textit{active} transformation, in which the coordinate axes are held fixed while the location of the observer in spacetime is boosted using the inverse of the transformation specified by \eqs{e10}{e11}.
That is, a spacetime point of the observer located at $(ct\,;\,\boldsymbol{\vec{x}})$ is transformed by the boost to
$(ct'\,;\,\boldsymbol{\vec{x}}^{\,\prime})$ using \eqs{e10}{e11} with $\boldsymbol{\vec{\zeta}}$ replaced by
$-\boldsymbol{\vec{\zeta}}$.  
Henceforth, all Lorentz transformations treated in this paper will correspond to active transformations.

The transformation that boosts the spacetime point $(ct\,;\,\boldsymbol{\vec{x}})$ to $(ct'\,;\,\boldsymbol{\vec{x}}^{\,\prime})$ is given by
\beq \label{e12}
\begin{pmatrix} ct' \\[8pt] x^{\prime i}\end{pmatrix} = \Lambda(\boldsymbol{\vec{\zeta}}\,,\,\boldsymbol{\vec{0}}) \,\begin{pmatrix} ct\\[8pt] x^j\end{pmatrix}\,,
\eeq
where the $4\times 4$ matrix $\Lambda(\boldsymbol{\vec{\zeta}}\,,\,\boldsymbol{\vec{0}})$ can be written in block matrix form as
\beq  \label{Lmatrixzeta}
\Lambda(\boldsymbol{\vec{\zeta}}\,,\,\boldsymbol{\vec{0}}) =
\left(\begin{array}{c|c} \cosh\zeta  & \,\,\, \displaystyle\frac{\zeta^j}{\zeta}\sinh\zeta \\[12pt] \hline \\[-10pt]
\phm \displaystyle\frac{\zeta^i}{\zeta}\sinh\zeta\phm & \,\,\, \delta^{ij}+\displaystyle\frac{\zeta^i\zeta^j}{|\boldsymbol{\vec{\zeta}}|^2}(\cosh\zeta-1) \end{array}\right)\,,
\eeq
after converting \eqs{e10}{e11} to an active transformation via $\boldsymbol{\vec{\zeta}}\to -\boldsymbol{\vec{\zeta}}$.  In \eq{Lmatrixzeta},
\beq
\delta^{ij}=\begin{cases} 1\,, & \text{if $i=j$}, \\ 0\,, & \text{if $i\neq j$},\end{cases}
\eeq
where the Latin indices 
$i,j\in\{1,2,3\}$ refer to the $x$, $y$, and $z$ components of the three-vector~$\boldsymbol{\vec{\zeta}}$, 
and there is an implicit sum over the repeated index $j$ on the right hand side of \eq{e12}.  

The matrix $\Lambda(\boldsymbol{\vec{\zeta}},\boldsymbol{\vec{0}})$ is sometimes inaccurately called the Lorentz transformation matrix.  In fact, this matrix represents a special type of Lorentz transformation consisting of a boost \textit{without} rotation [the latter is indicated by the second argument of $\Lambda(\boldsymbol{\vec{\zeta}}\,,\,\boldsymbol{\vec{0}})$].  Furthermore, note that
 $\Lambda(\boldsymbol{\vec{0}},\boldsymbol{\vec{0}})=\id_4$ is the $4\times 4$ identity matrix.   Any Lorentz transformation of the form $\Lambda(\boldsymbol{\vec{\zeta}}\,,\,\boldsymbol{\vec{0}})$
can be continuously deformed into the identity matrix by continuously shrinking the vector $\boldsymbol{\vec{\zeta}}$ to the zero vector.

Another example of a Lorentz transformation is a three-dimensional proper rotation 
of the vector $\boldsymbol{\vec{x}}$ into the vector $\boldsymbol{\vec{x}}^{\,\prime}=R\boldsymbol{\vec{x}}\,$ by an angle $\theta$, counterclockwise,  about a fixed axis~$\boldsymbol{\hat{n}}$, where~$R$ 
is a $3\times 3$ orthogonal matrix of unit determinant, and the time coordinate is not transformed. 
In this notation, $\boldsymbol{\hat{n}}=(n^1,n^2,n^3)$ is a unit vector (i.e.,
$\boldsymbol{\hat{n}\newcdot\hat{n}}=1$).   It is then convenient to define a three-vector quantity called the rotation vector,
\beq \label{vectheta}
\boldsymbol{\vec{\theta}}\equiv\theta\boldsymbol{\hat{n}}\,,
\eeq
where $0\leq\theta\leq\pi$.
In the case of a proper three-dimensional rotation,
the transformation of the spacetime point $(ct\,;\,\boldsymbol{\vec{x}})$ to $(ct'\,;\,\boldsymbol{\vec{x}}^{\,\prime})$ is given by
\beq
\begin{pmatrix} ct' \\[8pt] x^{\prime i}\end{pmatrix} = \Lambda(\boldsymbol{\vec{0}}\,,\,\boldsymbol{\vec{\theta}}) \,\begin{pmatrix} ct\\[8pt] x^j\end{pmatrix}\,,
\eeq
where the $4\times 4$ matrix $\Lambda(\boldsymbol{\vec{0}}\,,\,\boldsymbol{\vec{\theta}})$ can be written in block matrix form as
\beq  \label{Lmatrix2}
\Lambda(\boldsymbol{\vec{0}}\,,\,\boldsymbol{\vec{\theta}}) =
\left(\begin{array}{c|c}\phm 1 \phm  & \,\,\,0^j\\[3pt] \hline \\[-10pt]
\phm 0^i \phm& \,\,\,R^{ij}(\boldsymbol{\hat{n}},\theta)\end{array}\right)\,,
\eeq
where $0^j$ [$0^i$] are the components of the zero row [column] vector (with $i$, $j\in\{1,2,3\}$), and
\beq \label{Rijp}
R^{ij}(\boldsymbol{\hat{n}},\theta)=\delta^{ij}\cos\theta+n^in^j(1-\cos\theta)-\epsilon^{ijk}n^k\sin\theta\,.
\eeq
In \eq{Rijp}, the Levi--Civita symbol is defined by $\epsilon^{ijk}= +1$ [$-1$] when $ijk$ is an even [odd] permutation of 123, and $\epsilon^{ijk}= 0$ if any two of the indices coincide.
\Eq{Rijp} is known as Rodrigues' rotation formula (e.g., see Refs.~\cite{Marsden,Gallier}).   A clever proof of this formula is provided in Appendix~\ref{appA}.

\section{General Lorentz transformations}
\label{sec:L}

Consider a four-vector $v^\mu=(v^0\,;\boldsymbol{\vec{v}})$.  
Under an \textit{active}
Lorentz transformation, the spacetime components of the four-vector $v^\mu$ transform as
\beq \label{vp}
v^{\prime\mu} = \Lambda^\mu{}_\alpha v^\alpha\,,
\eeq
where the Greek indices such as $\mu$, $\alpha\in\{0,1,2,3\}$, and there is an implied sum over any repeated upper/lower index pair.  The quantities $\Lambda^\mu{}_\alpha$ can be viewed as the elements of a
$4\times 4$ real matrix, where $\mu$ labels the row and $\alpha$ labels the column.
In special relativity, the metric tensor (in a rectangular coordinate system) is given by the diagonal matrix. 
\beq \label{etamunu}
\eta_{\mu\nu}={\rm diag}(+1; -1,-1,-1)\,,
\eeq 
where the so-called mostly minus convention for the metric tensor has been chosen.

To construct a Lorentz-invariant scalar quantity that is unchanged under a Lorentz transformation, one only needs to combine tensors in such a way that
all upper/lower index pairs are summed over and no unsummed indices remain.   For example,
\beq \label{etavv}
\eta_{\mu\nu}v^{\prime\mu}v^{\prime\nu}=\eta_{\alpha\beta} v^\alpha v^\beta\,.
\eeq
Using \eqs{vp}{etavv}, it follows that
\beq
(\eta_{\mu\nu} \Lambda^\mu{}_\alpha \Lambda^\nu{}_\beta-\eta_{\alpha\beta}) v^\alpha v^\beta=0\,.
\eeq
Since the four-vector $v$ is arbitrary, it follows that
\beq \label{lambdarelation}
\Lambda^\mu{}_\alpha \eta_{\mu\nu}\Lambda^\nu{}_\beta=\eta_{\alpha\beta}\,.
\eeq
\Eq{lambdarelation} defines the most general Lorentz transformation matrix $\Lambda$.  The set of all such $4\times 4$ Lorentz transformation matrices is a group (under matrix multiplication) and is denoted by O$(1,3)$.   Here, the notation $(1,3)$ refers to the number of plus and minus signs in the metric tensor $\eta_{\mu\nu}$ [cf.~\eq{etamunu}].  In particular, 
O$(1,3)$ is a Lie group, appropriately called the Lorentz group (e.g., see Refs.~\cite{Sexl,Markoutsakis,Gallier}).

After taking the determinant of both sides of \eq{lambdarelation}, one obtains $(\det \Lambda)^2=1$.  Hence,
\beq \label{detlam}
\det \Lambda=\pm 1\,.
\eeq
Moreover, by setting $\alpha=\beta=0$ in \eq{lambdarelation} and summing over $\mu$ and $\nu$, one obtains
\beq \label{zerozero}
(\Lambda^0{}_0)^2=1+(\Lambda^1{}_0)^2+(\Lambda^2{}_0)^2+(\Lambda^3{}_0)^2\quad\Longrightarrow \quad (\Lambda^0{}_0)^2\geq 1\,.
\eeq

The Lie group SO$(1,3)$ is the group of \textit{proper} Lorentz transformation matrices that satisfy $\det \Lambda=+1$.
The elements of the subgroup of SO$(1,3)$ that also satisfy $\Lambda^0{}_0\geq +1$ are continuously connected to the identity element [the $4\times 4$ identity matrix, denoted by~$\boldsymbol{I}_4$] and constitute the set of proper orthochronous Lorentz transformations, which is often denoted by SO$_0(1,3)$. 
Three examples of Lorentz transformations that are not continuously connected to the identity are as follows 
\beqa
\!\!\!\!\!\!\Lambda_P\mspace{-1mu}=\mspace{-1mu}{\rm diag}(1,-1,-1,-1)\,,\qquad \!\! \Lambda_T\mspace{-1mu}=\mspace{-1mu}{\rm diag}(-1,1,1,1)\,,\qquad\!\!\Lambda_P\Lambda_T\mspace{-1mu}=\mspace{-1mu}{\rm diag}(-1,-1,-1,-1)\mspace{2mu}.
\eeqa
In particular, there is no way to continuously change the parameters of a proper ortho\-chronous Lorentz transformation to yield a Lorentz transformation with $\det\Lambda=-1$ and/or
$\Lambda^0{}_0\leq -1$ in light of \eqs{detlam}{zerozero}.

The complete list of Lorentz transformations is then given by
\beq
\bigl\{\Lambda\,, \Lambda_P\Lambda\,,\,\Lambda_T\Lambda\,,\,\Lambda_P\Lambda_T\Lambda\,|\, \Lambda\in {\rm SO}_0(1,3)\bigr\}.
\eeq
Consequently, to determine the explicit form of the most general Lorentz transformation, it suffices to consider the explicit form of the most general proper orthochronous Lorentz transformation.

The Lie algebra of the Lorentz group is obtained by considering an infinitesimal Lorentz transformation,
\beq \label{small}
\Lambda=\boldsymbol{I}_4+A\,,
\eeq
where $A$ is a $4\times 4$ matrix that depends on infinitesimal Lorentz group parameters.  In particular, terms that are quadratic or of higher order in the infinitesimal group parameters are neglected.   Inserting \eq{small} into \eq{lambdarelation}, and denoting $G={\rm diag}(+1,-1,-1,-1)$ to be the $4\times 4$ matrix whose matrix elements are $\eta_{\mu\nu}$, it follows that
\beq
\bigl(\boldsymbol{I}_4+A^{\T}\bigr)G\bigl(\boldsymbol{I}_4+A)=G\,.
\eeq
Keeping only terms up to linear order in the infinitesimal group parameters, we conclude that $A^{\T}G=-GA$ or equivalent (since $G$ is a diagonal matrix),
\beq \label{ATG}
(GA)^{\T}=-GA\,.
\eeq
That is, $GA$ is a $4\times 4$ real antisymmetric matrix.  Hence, the Lie algebra of the Lorentz group, henceforth denoted by $\mathfrak{so}(1,3)$, consists of all $4\times 4$ real matrices $A$ such that $GA$ is an antisymmetric matrix.  

To construct a proper orthochronous Lorentz transformation, one can choose any $4\times 4$ real matrix $A$ that satisfies \eq{ATG}, and consider a large positive integer $n$ such that $A/n$ is an infinitesimal quantity.  Then, a proper orthochronous Lorentz transformation is obtained by applying a sequence of $n$ infinitesimal Lorentz transformations in the limit as $n\to\infty$,
\beq \label{exponentiate}
\Lambda=\lim_{n\to\infty} \left(\boldsymbol{I}_4+\frac{A}{n}\right)^n=\exp A\,. 
\eeq
Note that $\Lambda$ is continuously connected to the identity matrix since one can continuously deform $A$ into the zero matrix.  Hence, it follows that $\Lambda\in{\rm SO}_0(1,3)$.
However, one can make a stronger statement: the exponential map, $\exp: \mathfrak{so}(1,3)\to {\rm SO}_0(1,3)$, is surjective.  A proof of this result can be found in Section 6.3 of Ref.~\cite{Gallier}. 
That is, the set of \textit{all} proper orthochronous Lorentz transformations consists of matrices of the form $\exp A$, where $GA$ is a $4\times 4$ real antisymmetric matrix.  

Let us first reconsider the two special cases examined in Section~\ref{intro}.  A matrix representation of an infinitesimal boost is obtained by evaluating \eq{Lmatrixzeta} to leading order in~$\zeta$,
\beq  \label{Lmatrixzetainf}
\Lambda(\boldsymbol{\vec{\zeta}}\,,\,\boldsymbol{\vec{0}}) \simeq
\left(\begin{array}{c|c}  \phantom{|} 1  \phantom{|} & \zeta^j \\[3pt] \hline \\[-10pt]
\phantom{|}\zeta^i  \phantom{|}& \,\,\, \delta^{ij} \end{array}\right)=\id_4-i\mathbold{{\vec\zeta}\newcdot}\boldsymbol{\vec k}+\mathcal{O}(|\boldsymbol{\vec{\zeta}}|^2)\,,
\eeq
where the three matrices $\boldsymbol{\vec{k}}=(k^1\,,\,k^2\,,\,k^3)$ are defined by
\beq
 k^1=i\begin{pmatrix} 0 &\phm 1 &\phm 0 &\phm 0 \\
1 & \phm 0 & \phm 0 &\phm 0 \\ 0 & \phm 0 & \phm 0 & \phm 0\\
0 & \phm 0 &\phm 0 & \phm 0\end{pmatrix}, \qquad
\!\!\! k^2=i\begin{pmatrix} 0 &\phm 0 &\phm 1 &\phm 0 \\
0 & \phm 0 & \phm 0 &\phm  0\\ 1  & \phm 0 & \phm 0 & \phm 0 \\
0 & \phm 0 &\phm  0 & \phm 0\end{pmatrix}, \qquad
\!\!\! k^3=i\begin{pmatrix} 0 &\phm 0&\phm 0 &\phm 1\\
0 & \phm 0 & \phm 0 &\phm 0\\ 0& \phm 0 & \phm 0 & \phm 0 \\
1 & \phm 0&\phm  0& \phm 0\end{pmatrix}.  \label{kmatrix}
\eeq
Similarly, a matrix representation of an infinitesimal rotation is obtained by evaluating \eqs{Lmatrix2}{Rijp} to leading order in $\theta$ (with $\theta^k\equiv\theta n^k$),
\beq \label{Lmatrix2inf}
\Lambda(\boldsymbol{\vec{0}}\,,\,\boldsymbol{\vec{\theta}}) \simeq
\left(\begin{array}{c|c} \phantom{|} \! 1  \phantom{|}   & \,\,\,0^j\\[2pt] \hline \\[-10pt]
 \phantom{||} 0^i \ \phantom{|} & \,\,\,\delta^{ij}-\epsilon^{ijk}\theta^k \end{array}\right)=\id_4-i\mathbold{{\vec\theta}\newcdot}\boldsymbol{\vec s}+\mathcal{O}(|\boldsymbol{\vec{\theta}}|^2)\,,
\eeq
where the three matrices $\boldsymbol{\vec{s}}=(s^1\,,\,s^2\,,\,s^3)$ are defined by
\beq
s^1= i\begin{pmatrix} 0 &\phm 0 &\phm 0 &\phm 0 \\
0 & \phm 0 & \phm 0 &\phm 0 \\ 0 & \phm 0 & \phm 0 & -1 \\
0 & \phm 0 &\phm 1 & \phm 0\end{pmatrix}, \qquad
\!\!\! s^2=i\begin{pmatrix} 0 &\phm 0 &\phm 0 &\phm 0 \\
0 & \phm 0 & \phm 0 &\phm 1 \\ 0 & \phm 0& \phm 0 & \phm 0 \\
0 & -1 &\phm 0 & \phm 0\end{pmatrix}, \qquad
\!\!\! s^3=i\begin{pmatrix} 0 &\phm 0 &\phm 0&\phm 0 \\
0 & \phm 0 & -1 &\phm 0 \\ 0 & \phm 1& \phm 0 & \phm 0 \\
0 & \phm 0 &\phm 0 & \phm 0\end{pmatrix}. \label{smatrix} 
\eeq
The six matrices $\boldsymbol{\vec{k}}=(k^1\,,\,k^2\,,\,k^3)$ and $\boldsymbol{\vec{s}}=(s^1\,,\,s^2\,,\,s^3)$ satisfy the following commutation relations:
\beq \label{sskk}
[s^i\,,\,s^j]=i\epsilon^{ij\ell}s^\ell\,,\qquad [k^i\,,\,k^j]=-i\epsilon^{ij\ell}s^\ell\,,\qquad[s^i\,,\,k^j]=i\epsilon^{ij\ell} k^\ell\,,
\eeq
where $i,j,\ell\in\{1,2,3\}$ and there is an implicit sum over the repeated index $\ell$.

Using \eqs{Lmatrixzetainf}{Lmatrix2inf}, it
follows that the matrix representation of a general infinitesimal Lorentz transformation, to linear order in the boost and rotation parameters, is given by
\beq \label{LamLaminf}
\Lambda(\boldsymbol{\vec{\zeta}}\,,\,\boldsymbol{\vec{\theta}})\simeq \Lambda(\boldsymbol{\vec{0}}\,,\,\boldsymbol{\vec{\theta}}) \Lambda(\boldsymbol{\vec{\zeta}}\,,\,\boldsymbol{\vec{0}})  \simeq
\left(\begin{array}{c|c}\phm 1 \phm & \,\,\, \zeta^j \\[3pt] \hline \\[-10pt]
\phm\zeta^i\phm & \,\,\,\delta^{ij}-\epsilon^{ijk}\theta^k\end{array}\right)\simeq \id_4-i\mathbold{{\vec\theta}\newcdot}\boldsymbol{\vec s}-i\mathbold{{\vec\zeta}\newcdot}\boldsymbol{\vec k}\,.
\eeq
Note that we also could have written $\Lambda(\boldsymbol{\vec{\zeta}}\,,\,\boldsymbol{\vec{\theta}})\simeq \Lambda(\boldsymbol{\vec{\zeta}}\,,\,\boldsymbol{\vec{0}}) \Lambda(\boldsymbol{\vec{0}}\,,\,
\boldsymbol{\vec{\theta}})$ in \eq{LamLaminf}, 
since the infinitesimal Lorentz transformations commute at linear order. 

In light of the remarks below \eq{exponentiate},
one can conclude that the most general proper orthochronous
Lorentz transformation matrix $\Lambda(\boldsymbol{\vec{\zeta}},\boldsymbol{\vec{\theta}})$ is a $4\times 4$ matrix given by 
\begin{equation} \label{lambda44}
\Lambda(\boldsymbol{\vec{\zeta}}\,,\,\boldsymbol{\vec{\theta}})=\exp\bigl(
-i\mathbold{{\vec\theta}\newcdot}\boldsymbol{\vec s}
-i\mathbold{{\vec\zeta}\newcdot}\boldsymbol{\vec k}\bigr)\,.
\end{equation}
Here, we follow the conventions of Refs.~\cite{Dreiner:2008tw,Dreiner:2023yus}.  
Note that in the notation of Ref.~\cite{Jackson}, $\boldsymbol{\vec{k}}=i\boldsymbol{\vec{K}}$ and
 $\boldsymbol{\vec{s}}=i\boldsymbol{\vec{S}}$, where the $4\times 4$ matrix representations 
 of $\boldsymbol{\vec{K}}$ and $\boldsymbol{\vec{S}}$ are given in eq.~(11.91) of Ref.~\cite{Jackson} and yield
 $\Lambda=\exp(\boldsymbol{\vec{\theta}\newcdot\vec{S}}+\boldsymbol{\vec{\zeta}\newcdot\vec{K}})$.  The argument of exp differs by an overall sign with eq.~(11.93) of Ref.~\cite{Jackson}, where a \textit{passive} Lorentz transformation is employed, which amounts to replacing $\{\boldsymbol{\vec\zeta},
\boldsymbol{\vec\theta}\}$ with $\{-\boldsymbol{\vec\zeta},
-\boldsymbol{\vec\theta}\}$.

\Eqss{kmatrix}{smatrix}{lambda44} imply that
\beq \label{Amatrix}
\Lambda(\boldsymbol{\vec{\zeta}}\,,\,\boldsymbol{\vec{\theta}})=\exp A\,,\quad \text{where}\quad A\equiv -i\mathbold{{\vec\theta}\newcdot}\boldsymbol{\vec s}
-i\mathbold{{\vec\zeta}\newcdot}\boldsymbol{\vec k}=\begin{pmatrix} 0 &\phm \zeta^1 &\phm \zeta^2 &\phm \zeta^3 \\
\zeta^1 & \phm\! 0 & -\theta^3 &\phm \theta^2 \\\zeta^2 & \phm\theta^3 & \phm\! 0 & -\theta^1 \\
\zeta^3 & -\theta^2 &\phm \theta^1 & \phm\! 0\end{pmatrix}.
\eeq
As anticipated in \eq{ATG}, $GA$ is the most general $4\times 4$ real antisymmetric matrix, which depends on six real independent parameters $\zeta^i$ and $\theta^i$
($i\in\{1,2,3\}$).
The $\{s^i\,,\,k^i\}$ satisfy the
commutation relations [\eq{sskk}] of the real Lie algebra $\mathfrak{so}(1,3)$.
As indicated in \eq{Amatrix}, $A$ is a real linear combination of the six Lie algebra generators $\{-is^i\,,\,-ik^i\}$ and thus constitutes a general element of $\mathfrak{so}(1,3)$.  
In Section~\ref{sec:fourbyfour}, we provide an explicit computation of $\exp A$.

Before moving on, we shall introduce a useful notation that assembles the matrices $\{s^i\,,\,k^i\}$ into six independent non-zero matrices, 
 $s^{\rho\lambda}=-s^{\lambda\rho}$ (with $\lambda$, $\rho\in\{0,1,2,3\}$) \mbox{such that}
 \beq \label{skdefs}
s^\ell \equiv \half\eps^{ij\ell}s^{ij}\,,\qquad 
k^i \equiv s^{0i}= -s^{i0}\,.
\eeq
Note that \eq{skdefs} implies that $s^{ij}=\epsilon^{ij\ell}s^\ell$, so that the six independent matrices can be taken to be $s^{ij}$ $(i<j)$ and $s^{0i}$ ($i,j\in\{1,2,3\}$).
The matrix elements of the $s^{\rho\lambda}$ are given by
\begin{equation} \label{explicitsmunu} 
(s^{\rho\lambda})^\mu{}_\nu=  i\bigl(\eta^{\rho\mu}\delta^\lambda_\nu-\eta^{\lambda\mu}\delta^\rho_\nu 
\bigr)\,,
\end{equation} 
where $\mu$ indicates the row and $\nu$ indicates the column of the corresponding matrix.   

Using \eq{skdefs}, one can check that \eq{explicitsmunu} is equivalent
to \eqs{kmatrix}{smatrix}.   In addition, 
the $\mathfrak{so}(1,3)$ commutation relations exhibited in \eq{sskk} now take the following form:
\beq \label{eq:comm-rels}
[s^{\alpha\beta},s^{\rho\lambda}] = i(\eta^{\beta\rho}\,s^{\alpha\lambda} -
\eta^{\alpha\rho}\,s^{\beta\lambda} - \eta^{\beta\lambda}\,s^{\alpha\rho} +
\eta^{\alpha\lambda}\,s^{\beta\rho} ).
\eeq
One can also assemble the boost and rotation parameters $\{\zeta^i\,,\,\theta^i\}$ into a second rank antisymmetric tensor $\theta^{\alpha\beta}$ by defining
\beq
\theta^{ij}\equiv\epsilon^{ij\ell}\theta^\ell\,,\qquad
\theta^{i0}=-\theta^{0i}\equiv\zeta^i\,.
\eeq
With this new notation, \eq{lambda44} can be rewritten as
\begin{equation} \label{lambda44again}
\Lambda(\boldsymbol{\vec{\zeta}}\,,\,\boldsymbol{\vec{\theta}})=\exp\left(-\half i\theta_{\rho\lambda}s^{\rho\lambda}\right)\,,
\end{equation}
where $\theta_{\rho\lambda}\equiv \eta_{\rho\alpha}\eta_{\lambda\beta}\theta^{\alpha\beta}$.
As usual, there is an implied sum over each pair of repeated upper/lower indices.

\section{An explicit evaluation of \texorpdfstring{$\Lambda(\boldsymbol{\vec{\zeta}}\,,\,\boldsymbol{\vec{\theta}})=\exp A$}{\uppercaseLambda(\textzeta\overarrow,\texttheta\overarrow)=exp A}}
\label{sec:fourbyfour}

We now proceed to evaluate  $\exp A$, where $A$ is given by \eq{Amatrix}.   First, we compute the characteristic polynomial of $A$,
\beq \label{charpoly}
p(x)\equiv 
\det(A-x\boldsymbol{I}_4)=x^4+\bigl(|\boldsymbol{\vec{\theta}}|^2-|\boldsymbol{\vec{\zeta}}|^2\bigr)x^2-(\boldsymbol{\vec{\theta}}\newcdot\boldsymbol{\vec{\zeta}})^2\equiv (x^2+a^2)(x^2-b^2)\,,
\eeq
where
\beq \label{characteristic}
a^2 b^2=(\boldsymbol{\vec{\theta}}\newcdot\boldsymbol{\vec{\zeta}})^2\,,\qquad\quad
a^2-b^2=|\boldsymbol{\vec{\theta}}|^2-|\boldsymbol{\vec{\zeta}}|^2\,.
\eeq
Solving \eq{characteristic} for $a^2$ and $b^2$ yields
\beqa
a^2&=&\frac12\biggl[|\boldsymbol{\vec{\theta}}|^2-|\boldsymbol{\vec{\zeta}}|^2+\sqrt{\bigl(|\boldsymbol{\vec{\theta}}|^2-|\boldsymbol{\vec{\zeta}}|^2\bigr)\lsup{2}+4(\boldsymbol{\vec\theta\newcdot\vec{\zeta}})^2}\,\biggr]\,,\label{a}\\[8pt]
b^2&=&\frac12\biggl[|\boldsymbol{\vec{\zeta}}|^2-|\boldsymbol{\vec{\theta}}|^2+\sqrt{\bigl(|\boldsymbol{\vec{\theta}}|^2-|\boldsymbol{\vec{\zeta}}|^2\bigr)\lsup{2}+4(\boldsymbol{\vec\theta\newcdot\vec{\zeta}})^2}\,\biggr]\,.\label{b}
\eeqa

Note that $a^2\geq 0$ and $b^2\geq 0$ so that $a,b\in\mathbb{R}$.   The individual signs of $a$ and $b$ are not determined, but none of the results that follow depend on these signs.
The eigenvalues of~$A$, denoted by $\lambda_i$ ($i=1,2,3,4$), are the solutions of $p(x)=0$, which are given by
\beq \label{lambdas}
\lambda_i=ia, -ia, b, -b\,.
\eeq
If $ab\neq 0$, then the four eigenvalues of $A$ [\eq{lambdas}] are distinct, which implies that $A$ is a diagonalizable matrix.

To evaluate $\exp A$ for a diagonalizable matrix $A$, we shall make use of a formula [\eq{fA} below] that is based on the Lagrange interpolating polynomial.  
Consider an  $n\times n$ 
matrix~$A$ with $n$ eigenvalues of which $m$ are distinct and denoted by $\lambda_i$ ($i=1,2,\ldots,m$).  The matrix $A$ is diagonalizable if and only if
\beq \label{minpoly}
\prod_{i=1}^m (A-\lambda_i\boldsymbol{I}_n)=0\,,
\eeq
where $\boldsymbol{I}_n$ is the $n\times n$ identify matrix.\footnote{See, e.g., Section 8.3.2 of Ref.~\cite{Carrell} or Section 7.11 of Ref.~\cite{Meyer}.}
Note that if $m=n$ (i.e., all $n$ eigenvalues are distinct), then $A$ is diagonalizable, since in this case
\eq{minpoly} is automatically satisfied due to the Cayley--Hamilton theorem. 

Any function of a diagonalizable matrix $A$ is given by the following formula:\footnote{See, 
e.g., eqs.~(7.3.6) and (7.3.11) of Ref.~\cite{Meyer}, eq.~(5.4.17) of Ref.~\cite{Mehta}, 
or Chapter V, Section~2.2 of Ref.~\cite{Gantmacher}.}
\beq \label{fA}
f(A)=\sum_{i=1}^m f(\lambda_i)K_i\,,\qquad
\text{where} \qquad 
K_i= \prod_{\substack{j=1 \\ j\neq i}}^m\frac{A-\lambda_j\boldsymbol{I}_n}{\lambda_i-\lambda_j}\,,
\eeq
if $2\leq m\leq n$ and $K_1\equiv \boldsymbol{I}_n$ if $m=1$.  Note that $\sum_{i=1}^m K_i = \boldsymbol{I}_n$.

Applying \eq{fA} to $f(A)=\exp A$, where $A$ is given by \eq{Amatrix}, under the assumption that $ab\neq 0$, it follows that
\beqa
\exp A&=&e^{ia}\left(\frac{A+ia\boldsymbol{I}_4}{2ia}\right)\!\left(\frac{A-b\boldsymbol{I}_4}{ia-b}\right)\!\left(\frac{A+b\boldsymbol{I}_4}{ia+b}\right)
+e^{-ia}\left(\frac{A-ia\boldsymbol{I}_4}{-2ia}\right)\!\left(\frac{A-b\boldsymbol{I}_4}{-ia-b}\right)\!\left(\frac{A+b\boldsymbol{I}_4}{-ia+b}\right)\nonumber \\[8pt]
&& \!\!\!\!\!\!\!\!\! +\, e^{b}\left(\frac{A-ia\boldsymbol{I}_4}{b-ia}\right)\!\left(\frac{A+ia\boldsymbol{I}_4}{b+ia}\right)\!\left(\frac{A+b\boldsymbol{I}_4}{2b}\right)
+e^{-b}\left(\frac{A-ia\boldsymbol{I}_4}{-b-ia}\right)\!\left(\frac{A+ia\boldsymbol{I}_4}{-b+ia}\right)\!\left(\frac{A-b\boldsymbol{I}_4}{-2b}\right).\nonumber \\
&& \phantom{line}
\eeqa
Simplifying the above expression yields,
\beq
\exp A=\frac{1}{a^2+b^2}\left\{(b^2\boldsymbol{I}_4-A^2)\left(A\frac{\sin a}{a}+\boldsymbol{I}_4\cos a\right)+(A^2+a^2\boldsymbol{I}_4)\left(A\frac{\sinh b}{b}+\boldsymbol{I}_4\cosh b\right)\right\}.
\eeq
Combining terms, we end up with
\beq \label{general}
\exp\begin{pmatrix} 0 &\phm \zeta^1 &\phm \zeta^2 &\phm \zeta^3 \\
\zeta^1 & \phm\! 0 & -\theta^3 &\phm \theta^2 \\\zeta^2 & \phm\theta^3 & \phm\! 0 & -\theta^1 \\
\zeta^3 & -\theta^2 &\phm \theta^1 & \phm\! 0\end{pmatrix}=\frac{1}{a^2+b^2}\biggl\{f_0(a,b)\boldsymbol{I}_4+
f_1(a,b)A+f_2(a,b)A^2+f_3(a,b)A^3\biggr\}\,,
\eeq
where $a$ and $b$ are defined in \eq{characteristic} and 
the $f_k(a,b)$ are given by
\beqa
f_0(a,b)&=&b^2\cos a+a^2\cosh b\,,\qquad\qquad f_1(a,b)=\frac{b^2}{a}\sin a+\frac{a^2}{b}\sinh b\,,\label{f01}\\
f_2(a,b)&=&\cosh b-\cos a\,,\qquad\qquad\qquad\, f_3(a,b)=\frac{\sinh b}{b}-\frac{\sin a}{a}\,,\label{f23}
\eeqa
in agreement with the results previously obtained in Refs.~\cite{ZR,Geyer,DM,AR}.

The matrix $A$ and its powers can be conveniently written in block matrix form,
\beqa 
A&=&\left(\begin{array}{c|c}  \phm 0 \phm & \quad \zeta^j  \\[2pt] \hline \\[-10pt]
\phm \zeta^i \phm&\quad  -\epsilon^{ijk}\theta^k\end{array}\right)\,,\qquad\quad
 A^2=\left(\begin{array}{c|c} |\boldsymbol{\vec{\zeta}}|^2 & \quad  \epsilon^{jk\ell}\zeta^k\theta^\ell  \\[2pt] \hline \\[-10pt]
\phm   -\epsilon^{ik\ell}\zeta^k\theta^\ell \phm & \quad \zeta^i\zeta^j+\theta^i\theta^j-\delta^{ij}|\boldsymbol{\vec{\theta}}|^2\end{array}\right)\,,\label{IAAA}
\eeqa
and
\beq
   A^3 =\left(\begin{array}{c|c} 0 & \quad \bigl(|\boldsymbol{\vec{\zeta}}|^2-|\boldsymbol{\vec{\theta}}|^2\bigr)\zeta^j+(\boldsymbol{\vec{\theta}\newcdot\vec{\zeta}})\theta^j  \\[2pt] \hline \\[-10pt]
   \phm\bigl(|\boldsymbol{\vec{\zeta}}|^2-|\boldsymbol{\vec{\theta}}|^2\bigr)\zeta^i+(\boldsymbol{\vec{\theta}\newcdot\vec{\zeta}})\theta^i \phm & \quad (\epsilon^{jk\ell}\zeta^i-\epsilon^{ik\ell}\zeta^j)\zeta^k\theta^\ell+
   \epsilon^{ijk}\theta^k|\boldsymbol{\vec{\theta}}|^2
\end{array}\right).\label{Acubed}
 \eeq
 The $ij$ element of $A^3$ can be simplified by noting that the $ij$ element of any $3\times 3$ antisymmetric matrix must be of the form $\epsilon^{ijk}C^k$ (after summing over the repeated index $k$).  Thus,
\beq
 (\epsilon^{jk\ell}\zeta^i-\epsilon^{ik\ell}\zeta^j)\zeta^k\theta^\ell
=\epsilon^{ijk}C^k\,.
\eeq
Multiplying the above equation by $\epsilon^{ijm}$ and summing over $i$ and $j$ yields
\beq
(\delta^{i\ell}\delta^{km}-\delta^{ik}\delta^{\ell m})\zeta^i\zeta^k\theta^\ell-(\delta^{jk}\delta^{\ell m}-\delta^{j\ell}\delta^{km})\zeta^j\zeta^k\theta^\ell=2\delta^{km}C^k\,.
\eeq
It follows that
$
C^m=(\boldsymbol{\vec{\theta}\newcdot\vec{\zeta}})\zeta^m-|\boldsymbol{\vec{\zeta}}|^2\theta^m\,.
$
That is, we have derived the identity
\beq \label{id}
 (\epsilon^{jk\ell}\zeta^i-\epsilon^{ik\ell}\zeta^j)\zeta^k\theta^\ell
=\epsilon^{ijk}\bigl[(\boldsymbol{\vec{\theta}\newcdot\vec{\zeta}})\zeta^k-|\boldsymbol{\vec{\zeta}}|^2\theta^k\bigr]\,.
   \eeq
Thus, the matrix $A^3$ [\eq{Acubed}] can be rewritten in a more convenient form,
\beq \label{A3}
 A^3 = \left(\begin{array}{c|c}  0 & \quad \bigl(|\boldsymbol{\vec{\zeta}}|^2-|\boldsymbol{\vec{\theta}}|^2\bigr)\zeta^j+(\boldsymbol{\vec{\theta}\newcdot\vec{\zeta}})\theta^j  \\[2pt] \hline \\[-10pt]
  \phm \bigl(|\boldsymbol{\vec{\zeta}}|^2-|\boldsymbol{\vec{\theta}}|^2\bigr)\zeta^i+(\boldsymbol{\vec{\theta}\newcdot\vec{\zeta}})\theta^i \phm & \quad \epsilon^{ijk}\bigl[(\boldsymbol{\vec{\theta}\newcdot\vec{\zeta}})\zeta^k-\bigl(|\boldsymbol{\vec{\zeta}}|^2-|\boldsymbol{\vec{\theta}}|^2\bigr)\theta^k\bigr]
 \end{array}\right).
   \eeq
   
Consider separately the case of $ab=0$.   The eigenvalues given in \eq{lambdas} are no longer distinct.   If $a=0$ and $b\neq 0$, then the matrix $A$ is diagonalizable since $A$ satisfies \eq{minpoly}, i.e., $A(A^2-b^2\boldsymbol{I}_4)=0$.   In particular, if $a=0$ then \eq{characteristic} implies that $\boldsymbol{\vec{\theta}}\newcdot\boldsymbol{\vec{\zeta}}=0$ and 
$b^2=|\boldsymbol{\vec{\zeta}}|^2-|\boldsymbol{\vec{\theta}}|^2$.  Plugging these results into \eqs{IAAA}{A3} yields $A^3-b^2A=0$.  Consequently, one can make use of \eq{fA} with $m=3$ to obtain
\beqa 
\begin{array}{lll}
\exp A&=&\left(\displaystyle\frac{A-b\boldsymbol{I}_4}{-b}\right)\left(\displaystyle\frac{A+b\boldsymbol{I}_4}{b}\right)+e^b\left(\displaystyle\frac{A}{b}\right)\left(\displaystyle\frac{A+b\boldsymbol{I}_4}{2b}\right)
+e^{-b}\left(\displaystyle\frac{A}{-b}\right)\left(\displaystyle\frac{A-b\boldsymbol{I}_4}{-2b}\right)  \\[13pt]
&=&\boldsymbol{I}_4+\displaystyle\frac{\sinh b}{b}A+\left(\displaystyle\frac{\cosh b-1}{b^2}\right)A^2\,,\qquad \text{for $a=0$}. \label{Aazero}
\end{array}
\eeqa
One can check that \eq{Aazero} coincides with the $a\to 0$ limit of \eqst{general}{f23} after making use of $A^3=b^2 A$.  

Likewise, if $b=0$ and $a\neq 0$, then the matrix $A$ is diagonalizable since $A$ satisfies \eq{minpoly}, i.e., $A(A^2+a^2\boldsymbol{I}_4)=0$.   In particular, if $b=0$, then \eq{characteristic} implies that $\boldsymbol{\vec{\theta}}\newcdot\boldsymbol{\vec{\zeta}}=0$ and 
$a^2=|\boldsymbol{\vec{\theta}}|^2-|\boldsymbol{\vec{\zeta}}|^2$.  Plugging these results into \eqs{IAAA}{A3} yields $A^3+a^2A=0$.  Consequently, one can make use of \eq{fA} with $m=3$ to obtain
\beqa 
\begin{array}{lll}
\exp A &=& \left(\displaystyle\frac{A-ia\boldsymbol{I}_4}{-ia}\right)\left(\displaystyle\frac{A+ia\boldsymbol{I}_4}{ia}\right)+e^{ia}\left(\displaystyle\frac{A}{ia}\right)\left(\displaystyle\frac{A+ia\boldsymbol{I}_4}{2ia}\right)
+e^{-ia}\left(\displaystyle\frac{A}{-ia}\right)\left(\displaystyle\frac{A-ia\boldsymbol{I}_4}{-2ia}\right)  \phantom{xx} \\[13pt]
&=&\boldsymbol{I}_4+\displaystyle\frac{\sin a}{a}A+\left(\displaystyle\frac{1-\cos a}{a^2}\right)A^2\,,\qquad \text{for $b=0$}. \label{Abzero}
\end{array}
\eeqa
One can check that \eq{Abzero} coincides with the $b\to 0$ limit of \eqst{general}{f23} after making use of $A^3=-a^2 A$.

Finally, in the case of $a=b=0$, \eq{characteristic} yields $\boldsymbol{\vec{\theta}}\newcdot\boldsymbol{\vec{\zeta}}=0$ and 
$|\boldsymbol{\vec{\zeta}}|^2=|\boldsymbol{\vec{\theta}}|^2$.   Using \eq{A3}, it then follows that $A^3=0$.  Thus, the Taylor series of the exponential terminates and one obtains
\beq \label{Aabzero}
\exp A=1+A+\half A^2\,, \qquad \text{for $a=b=0$}.
\eeq
Although one cannot directly employ \eq{fA} in this final case (since $A$ is no longer diagonalizable), one can still recover \eq{Aabzero} either by taking the $b\to 0$ limit of \eq{Aazero} or the $a\to 0$ limit of \eq{Abzero}.

It is instructive to check the two limiting cases exhibited in Section~\ref{intro}.  First, if $\boldsymbol{\vec{\theta}}=\boldsymbol{\vec{0}}$, then $a=0$ and $b= |\boldsymbol{\vec{\zeta}}|\equiv \zeta$.  It then follows that
\beq \label{Acasei}
A= \left(\begin{array}{c|c} 0 & \,\,\, \zeta^j \\[2pt] \hline \\[-10pt] \phm\zeta^i \phm & \,\,\,\boldsymbol{0}^{ij} \end{array}\right)\,,\qquad\quad
A^2= \left(\begin{array}{c|c}  |\boldsymbol{\vec{\zeta}}|^2 & \phm \boldsymbol{\vec{0}}  \\[2pt] \hline \\[-10pt] 
\phantom{|} \boldsymbol{\vec{0}}\phantom{|}   & \phm \zeta^i\zeta^j  \end{array}\right)\,,
\eeq  
where $\boldsymbol{0}^{ij}$ is a $3\times 3$ matrix of zeros.  Using \eqs{Aazero}{Acasei}, we obtain
\beq
\Lambda(\boldsymbol{\vec{\zeta}}\,,\,\boldsymbol{\vec{0}}) =\left(\begin{array}{c|c} \cosh\zeta  & \,\,\, \displaystyle\frac{\zeta^j}{\zeta}\sinh\zeta \\[12pt] \hline \\[-10pt]
 \phm\displaystyle\frac{\zeta^i}{\zeta}\sinh\zeta\phm & \,\,\, \delta^{ij}+\displaystyle\frac{\zeta^i\zeta^j}{|\boldsymbol{\vec{\zeta}}|^2}(\cosh\zeta-1) \end{array}\right), \label{AA}
\eeq
in agreement with \eq{Lmatrixzeta}.  

Second, if $\boldsymbol{\vec{\zeta}}=\boldsymbol{\vec{0}}$, then $a= |\boldsymbol{\vec{\theta}}|\equiv\theta$ and $b=0$.  It follows that
\beq \label{AAA}
A=\left(\begin{array}{c|c} 0& \boldsymbol{\vec{0}}   \\[2pt] \hline \\[-10pt]
\phantom{|} \boldsymbol{\vec{0}}\phantom{|}   &  \phantom{|} -\epsilon^{ijk}\theta^k \end{array}\right)\,,\qquad\quad
 A^2=\left(\begin{array}{c|c} 0  & \boldsymbol{\vec{0}}  \phm  \\[2pt] \hline \\[-10pt]
\phantom{|} \boldsymbol{\vec{0}}\phantom{|}  & \phm \theta^i\theta^j-\delta^{ij}|\boldsymbol{\vec{\theta}}|^2 \end{array}\right)\,.
 \eeq
Using \eqs{Abzero}{AAA}, we end up with
\beqa  
\Lambda(\boldsymbol{\vec{0}}\,,\,\boldsymbol{\vec{\theta}}) =\left(\begin{array}{c|c} \phantom{xx} 1 \phantom{xx} &  \boldsymbol{\vec{0}}\phantom{|} \\[2pt] \hline \\[-10pt]
\phantom{xx} \boldsymbol{\vec{0}}\phantom{xx}  &\phantom{|} \delta^{ij}\cos\theta+n^i n^j(1-\cos\theta)-\epsilon^{ijk}n^k\sin\theta \end{array}\right), 
\label{AAArod}
\eeqa
after identifying $\theta^i=\theta n^i$.  We have thus recovered \eq{Lmatrix2} and Rodrigues' rotation formula [\eq{Rijp}].

A final limiting case of interest is the most general orthochronous Lorentz transformation in $2+1$ spacetime dimensions.  In this case, we can choose 
$\boldsymbol{\vec{\theta}}=\theta\boldsymbol{\hat{z}}$ and $\boldsymbol{\vec{\zeta}}=\zeta^1\boldsymbol{\hat{x}}+\zeta^2\boldsymbol{\hat{y}}$, which implies that 
$ab=0$ [cf.~\eq{characteristic}].  
Without loss of generality, one can take $b=0$ and $a^2= \theta^2-|\boldsymbol{\vec{\zeta}}|^2$, where $\theta^2$ is the square of the rotation angle $\theta$
(in two space dimensions,
there is no danger in confusing $\theta^2$ with the second component of the vector $\boldsymbol{\vec{\theta}}=\theta\boldsymbol{\hat{z}}$).   Hence, 
\eq{Abzero} yields
\beq \label{Lam2d1}
\exp\begin{pmatrix} 0 &\phm \zeta^1 &\phm \zeta^2 \\
\zeta^1 & \phm\! 0 & -\theta  \\\zeta^2 & \phm\theta & \phm\! 0 \end{pmatrix}
=\boldsymbol{I}_3+\left(\frac{\sin\sqrt{\theta^2-|\boldsymbol{\vec{\zeta}}|^2}}{\sqrt{\theta^2-|\boldsymbol{\vec{\zeta}}|^2}}\,\right)A
+\left(\frac{1-\cos\sqrt{\theta^2-|\boldsymbol{\vec{\zeta}}|^2}}{\theta^2-|\boldsymbol{\vec{\zeta}}|^2}\,\right)A^2\,,
\eeq
where the $3\times 3$ matrices $A$ and $A^2$ are given in block-diagonal form by
\beq \label{Lam2d2}
A=\left(\begin{array}{c|c}  \phm 0 \phm & \quad \zeta^j  \\[2pt] \hline \\[-10pt]
\phm \zeta^i \phm&\quad  -\theta\epsilon^{ij}\end{array}\right)\,,\qquad\quad
 A^2=\left(\begin{array}{c|c} |\boldsymbol{\vec{\zeta}}|^2 & \quad  -\theta\epsilon^{kj}\zeta^k \\[2pt] \hline \\[-10pt]
\phm   -\theta\epsilon^{ik}\zeta^k\phm & \quad \zeta^i\zeta^j-\theta^2\delta^{ij}\end{array}\right)\,,
\eeq 
with $i$, $j\in\{1,2\}$ (and an implied sum over $k=1,2$), $\epsilon^{12}=-\epsilon^{21}=1$, and $\epsilon^{11}=\epsilon^{22}=0$.

\section{An explicit evaluation of \texorpdfstring{$\Lambda^\mu{}_\nu=\half\Tr\bigl(M^\dagger\sigmabar^\mu M\sigma_\nu\bigr)$}{\uppercaseLambda
\superu\subv=\textonehalf Tr(M\textCross\textsigma\overbar\superu M\textsigma\subv)}}
\label{sec:twobytwo}

In Section~\ref{sec:L}, we remarked that a
general element of the Lie algebra $\mathfrak{so}(1,3)$ is a real linear combination of the six generators $\{-is^i\,,\,-ik^i\}$.   In particular, the matrix $A$ defined in \eq{Amatrix}
provides a four-dimensional matrix representation of $\mathfrak{so}(1,3)$.  
The corresponding $4\times 4$ matrix that represents a general element of the proper orthochronous Lorentz group, SO$_0$(1,3), is then obtained by exponentiation, $\Lambda(\boldsymbol{\vec{\zeta}},\boldsymbol{\vec{\theta}})=\exp A$.
In this section, we will take advantage of the existence of a two-dimensional matrix representation of $\mathfrak{so}(1,3)$.  It is noteworthy that by exponentiating this two-dimensional representation, one obtains a two-dimensional matrix representation of the group of complex $2\times 2$ matrices with unit determinant, which defines the Lie group ${\rm SL}(2,\mathbb{C})$. Thus, the two-dimensional matrix representation of ${\rm SL}(2,\mathbb{C})$ provides representation matrices $M$ [defined in \mbox{\eq{emdef}} below] for the elements of SO$_0$(1,3).  However, in this case, the $2\times 2$ matrices $M$ and $-M$ of ${\rm SL}(2,\mathbb{C})$ represent the \textit{same} element of SO$_0$(1,3)
[cf.~\eq{L4}].  

For example, consider the general element of the two-dimensional representation of ${\rm SL}(2,\mathbb{C})$ that is given by
\beq \label{emdef}
M=\exp\left(-\half i\boldsymbol{\vec{\theta}\newcdot\vec{\sigma}}-\half\boldsymbol{\vec{\zeta}\newcdot\vec{\sigma}}\right)\,,
\eeq
where $\boldsymbol{\vec{\zeta}}$ and $\boldsymbol{\vec{\theta}}$ are the boost and rotation vectors that parametrize an element of the proper orthochronous Lorentz group and 
$\boldsymbol{\vec{\sigma}}=(\sigma^1\,,\,\sigma^2\,,\,\sigma^3)$ are the three Pauli matrices assembled into a vector whose components are the $2\times 2$ matrices,
\beq \label{paulimatrices}
\sigma^1 = \begin{pmatrix} 0&\,\,\,\phm 1\\ 1&\,\,\,\phm 0\end{pmatrix}\,,\qquad
\sigma^2 = \begin{pmatrix} 0&\,\,\,-i\\ i&\,\,\,\phm0\end{pmatrix}\,,\qquad
\sigma^3 = \begin{pmatrix} 1&\,\,\,\phm 0\\ 0&\,\,\,-1\end{pmatrix}\,.
\eeq
It is convenient to define a fourth Pauli matrix, $\sigma^0=\id_2$,
where $\id_2$ is the $2\times 2$ identity matrix.  We can then define the four Pauli matrices in a unified notation.  Following the notation of 
Refs.~\cite{Dreiner:2008tw,Dreiner:2023yus}, we define
\beq \label{sigmadefs}
\sigma^\mu=(\boldsymbol{I}_2\,;\,\mathbold{\vec\sigma})\,,\qquad\quad
\sigmabar^\mu=(\boldsymbol{I}_2\,;\,-\mathbold{\vec\sigma})\,,
\eeq
where $\mu\in\{0,1,2,3\}$.
Note that these sigma matrices have been defined
with an upper (contravariant) index.  They are related to 
sigma matrices with a lower (covariant) index in the usual way:
\beq
\sigma_\mu=\eta_{\mu\nu}\sigma^\nu=
(\boldsymbol{I}_2\,;\,
- \mathbold{\vec\sigma})
\,,\qquad\qquad
\sigmabar_\mu=\eta_{\mu\nu}\sigmabar^\nu=
(\boldsymbol{I}_2\,;\,
 \mathbold{\vec\sigma})\,.
\eeq
However, the use of the spacetime indices $\mu$ and $\nu$ is slightly deceptive since
the sigma matrices defined above are \textit{fixed} matrices that do not change under a Lorentz transformation.

It is also convenient to introduce the set of $2\times 2$ matrices,
\beq \label{sigmamunu}
\sigma^{\mu\nu}=-\sigma^{\nu\mu}\equiv \tfrac14 i\bigl(\sigma^\mu\sigmabar^\nu-\sigma^\nu\sigmabar^\mu\bigr)\,.
\eeq
One can then rewrite \eq{emdef} in the following form that is reminiscent of \eq{lambda44again},
\beq \label{M}
M=\exp\left(-\half i\theta_{\mu\nu}\sigma^{\mu\nu}\right)\,.
\eeq
That is, the six independent $-i\sigma^{\mu\nu}$ matrices are generators of the 
Lie algebra of ${\rm SL}(2,\mathbb{C})$, henceforth denoted by $\mathfrak{sl}(2,\mathbb{C})$.
It is straightforward to check that the $2\times 2$ matrices $\sigma^{\mu\nu}$ possess the same commutation relations as the $4\times 4$ matrices $s^{\mu\nu}$ [cf.~\eq{eq:comm-rels}], which establishes the isomorphism $\mathfrak{so}(1,3)\simeq \mathfrak{sl}(2,\mathbb{C})$.
\clearpage

Under an active Lorentz transformation, a two-component spinor $\chi_\alpha$ (where $\alpha\in\{1,2\}$) transforms as
\beq \label{eq:lor-lower-undot}
\chi^\prime_\alpha= {M_\alpha}^\beta \chi_\beta, \qquad
\alpha,\beta \in\{1,2\}.
\eeq
Suppose that $\chi$ and $\eta$ are two-component spinors and consider the spinor product $\eta^\dagger\sigmabar^\mu\chi$.  Under a Lorentz transformation,
\beq \label{under}
\eta^\dagger\sigmabar^\mu\chi\longrightarrow (M\eta)^\dagger\sigmabar^\mu (M\chi)=\eta^\dagger(M^\dagger\sigmabar^\mu M)\chi\,.
\eeq
We assert that the quantity $\eta^\dagger\sigmabar^\mu\chi$ transforms as a Lorentz four-vector,
\beq \label{assert}
\eta^\dagger\sigmabar^\mu\chi\longrightarrow \Lambda^\mu{}_\nu \,\eta^\dagger\sigmabar^\nu\chi\,.
\eeq
The standard proof of this assertion based on the analysis of infinitesimal Lorentz transformations is given in Appendix~\ref{appB}.  
(See also Appendix~\ref{appC}, where the corresponding result is obtained by employing the four-component spinor formalism.)
\Eqs{under}{assert} imply that the following identity must be satisfied:
\beq \label{Lam}
M^\dagger \sigmabar^\mu M=\Lambda^\mu{}_\nu\sigmabar^\nu\,.
\eeq
Multiplying \eq{Lam} on the right by $\sigma_\rho$ and using $\Tr(\sigmabar^\nu\sigma_\rho)=2\delta^\nu_{\rho}$, it follows that
\beq \label{L4}
\Lambda^\mu{}_\nu=\half\Tr\bigl(M^\dagger\sigmabar^\mu M\sigma_\nu\bigr)\,.
\eeq

It is now convenient to introduce the complex vector, $\boldsymbol{\vec{z}}\equiv \boldsymbol{\vec{\zeta}}+i\boldsymbol{\vec{\theta}}$,
and the associated quantity,
\beq \label{Deltadef}
\Delta\equiv \bigl(\boldsymbol{\vec{z}\newcdot\vec{z}}\bigr)^{1/2}
=\bigl(|\boldsymbol{\vec{\zeta}}|^2-|\boldsymbol{\vec{\theta}}|^2+2i\boldsymbol{\vec{\theta}\newcdot\vec{\zeta}}\,\bigr)^{1/2}\,.
\eeq
One can now evaluate the matrix exponential $M=\exp\bigl(-\half\boldsymbol{\vec{z}\newcdot\vec{\sigma}}\bigr)$  [cf.~\eq{emdef}] by making use of \eq{fA} if $\Delta\neq 0$.   The corresponding 
eigenvalues of $-\half\boldsymbol{\vec{z}\newcdot\vec{\sigma}}$ are $\lambda=\pm\half\Delta$.   Hence,
\beqa 
\begin{array}{lll}
M=\exp\left(-\half\boldsymbol{\vec{z}\newcdot\vec{\sigma}}\right)&=& e^{\Delta/2}\left(\displaystyle\frac{\id_2\Delta-\boldsymbol{\vec{z}\newcdot\vec{\sigma}}}{2\Delta}\right)
+e^{-\Delta/2}\left(\displaystyle\frac{\id_2\Delta+\boldsymbol{\vec{z}\newcdot\vec{\sigma}}}{2\Delta}\right)  \\[10pt]
&=&\boldsymbol{I}_2\cosh\bigl(\half\Delta\bigr)-\boldsymbol{\vec{z}\newcdot\vec{\sigma}}\,\displaystyle\frac{\sinh\bigl(\half\Delta\bigr)}{\Delta}\,.\label{M1}
\end{array}
\eeqa
Note that the limit as $\Delta\to 0$ is continuous and yields $M=\boldsymbol{I}_2-\half \boldsymbol{\vec{z}\newcdot\vec{\sigma}}$.

Since the Pauli matrices are hermitian,
\beq \label{M2}
M^\dagger=\exp\left(-\half\boldsymbol{\vec{z}^{\,*}\newcdot\vec{\sigma}}\right)=\boldsymbol{I}_2\cosh\bigl(\half\Delta^*\bigr)-\boldsymbol{\vec{z}^{\,*}\newcdot\vec{\sigma}}\,\frac{\sinh\bigl(\half\Delta^*\bigr)}{\Delta^*}\,.
\eeq
We shall evaluate $\Lambda^\mu{}_{\nu}$ in four separate cases depending whether the spacetime index is 0 or $i\in\{1,2,3\}$.   In particular, using block matrix notation,
\eq{L4} yields
\beq  \label{Lmatrix3}
\Lambda(\boldsymbol{\vec{\zeta}}\,,\,\boldsymbol{\vec{\theta}}) =
\left(\begin{array}{c|c}\Lambda^0{}_0  & \,\Lambda^0{}_j\\[3pt] \hline \\[-10pt]
\Lambda^i{}_0 & \,\Lambda^i{}_j\end{array}\right)=
\frac12\left(\begin{array}{c|c}\Tr(M^\dagger M) & \,\,\, -\Tr (M\sigma^j M^\dagger)\\[3pt] \hline \\[-10pt]
-\Tr (M^\dagger\sigma^i M)\ & \,\,\,\Tr(M^\dagger \sigma^i M\sigma^j)\end{array}\right),
\eeq
where we have used $\sigma_j=-\sigma^j$ to obtain the final matrix expression above.
\clearpage

Plugging \eqs{M1}{M2} into \eq{L4} and evaluating the traces, 
\beqa
\Tr(\sigma^i\sigma^j) &=& 2\delta^{ij}\,,\\
\Tr(\sigma^i\sigma^j\sigma^k) &=& 2i\epsilon^{ijk}\,,\\
\Tr(\sigma^i\sigma^j\sigma^k\sigma^\ell) &=& 2(\delta^{ij}\delta^{k\ell}-\delta^{ik}\delta^{j\ell}+\delta^{i\ell}\delta^{jk})\,,
\eeqa
we end up with the following expressions:
\beqa
\Lambda^0{}_0&=&|\cosh\bigl(\half\Delta\bigr)|\lsup{2}+\left|\frac{\sinh\bigl(\half\Delta\bigr)}{\Delta}\right|\lsup{2}\bigl(|\boldsymbol{\vec{\zeta}}|^2+|\boldsymbol{\vec{\theta}}|^2\bigr)\,, \label{LT1}\\
\Lambda^0{}_j &=& \left(\frac{\cosh\bigl(\half\Delta^*\bigr)\sinh\bigl(\half\Delta\bigr)}{\Delta}\,z^j +{\rm c.c.}\right)+i\,\left|\frac{\sinh\bigl(\half\Delta\bigr)}{\Delta}\right|\lsup{2}\epsilon^{jk\ell}z^{k}z^{*\ell}\,, \label{LT2}\\
\Lambda^i{}_0 &=& \left(\frac{\cosh\bigl(\half\Delta^*\bigr)\sinh\bigl(\half\Delta\bigr)}{\Delta}\,z^i +{\rm c.c.}\right)+i\,\left|\frac{\sinh\bigl(\half\Delta\bigr)}{\Delta}\right|\lsup{2}\epsilon^{ik\ell}z^{*k}z^\ell\,, \label{LT3}\\
\Lambda^i{}_j&=&\biggr\{|\cosh\bigl(\half\Delta\bigr)|^{2}-\left|\frac{\sinh\bigl(\half\Delta\bigr)}{\Delta}\right|\lsup{2}\bigl(|\boldsymbol{\vec{\zeta}}|^2+|\boldsymbol{\vec{\theta}}|^2\bigr)\biggl\}\delta^{ij}+(z^{*i}z^j+z^i z^{*j})\left|\frac{\sinh\bigl(\half\Delta\bigr)}{\Delta}\right|\lsup{2}
\nonumber  \\
&& \qquad\qquad
+\left(\frac{i\sinh\bigl(\half\Delta\bigr)\cosh\bigl(\half\Delta^*\bigr)}{\Delta}\epsilon^{ijk}z^k+{\rm c.c.}\right)\,,\label{LT4}
\eeqa
where ${\rm c.c.}$ means the complex conjugate of the previous term and $\Delta$ is defined in\linebreak   \mbox{\eq{Deltadef}}.
Note that since $\Delta$ is a complex quantity,
 $|\!\cosh\bigl(\half\Delta\bigr)|^2=\cosh\bigl(\half\Delta\bigr)\cosh\bigl(\half\Delta^*\bigr)$
 and  $|\!\sinh\bigl(\half\Delta\bigr)/\Delta|^2=\sinh\bigl(\half\Delta\bigr)\sinh\bigl(\half\Delta^*\bigr)/|\Delta|^2$ in \eqst{LT1}{LT4}.

We can check the results of \eqst{LT1}{LT4} in three special cases.  First, consider the case of a pure boost, where $\boldsymbol{\vec{\theta}}=\boldsymbol{\vec{0}}$.  Then, $\boldsymbol{\vec{z}}=\boldsymbol{\vec{z}^{\,*}}= \boldsymbol{\vec{\zeta}}$ and $\Delta=|\boldsymbol{\vec{\zeta}}|\equiv\zeta$.  Plugging these values into \eqst{LT1}{LT4} yields the following block matrix form:
\beq \label{boost}
\Lambda(\boldsymbol{\vec{\zeta}}\,,\,\boldsymbol{\vec{0}}) =\left(\begin{array}{c|c} \cosh\zeta  & \,\,\, \displaystyle\frac{\zeta^j}{\zeta}\sinh\zeta \\[12pt] \hline \\[-10pt]
 \phm\displaystyle\frac{\zeta^i}{\zeta}\sinh\zeta\phm & \,\,\, \delta^{ij}+\displaystyle\frac{\zeta^i\zeta^j}{|\boldsymbol{\vec{\zeta}}|^2}(\cosh\zeta-1) \end{array}\right),
 \eeq
 which again reproduces the result of \eq{Lmatrixzeta}.

 Second, consider the case of $\boldsymbol{\vec{\zeta}}=\boldsymbol{\vec{0}}$.  Then, $\boldsymbol{\vec{z}}=-
\boldsymbol{\vec{z}^{\,*}}=i\boldsymbol{\vec{\theta}}$ and $\Delta=i\theta$.   Plugging these values into \eqst{LT1}{LT4} and writing $\theta^i=\theta n^i$ yields
\beq \label{LRod}
\Lambda(\boldsymbol{\vec{0}}\,,\,\boldsymbol{\vec{\theta}}) =\left(\begin{array}{c|c}  \mspace{2mu}1 \phantom{|} & 0^j \\[2pt] \hline \\[-10pt]
\phantom{|}0^i \phantom{|}  &\phantom{|} \delta^{ij}\cos\theta+n^i n^j(1-\cos\theta)-\epsilon^{ijk}n^k\sin\theta \end{array}\right).
\eeq
Once again, we have recovered \eq{Lmatrix2} and Rodrigues' rotation formula [\mbox{\eq{Rijp}}].

Third, one can check that \eqst{LT1}{LT4} reduce to the most general orthochronous Lorentz transformation in $2+1$ spacetime dimensions for $i,j\in\{1,2\}$ if
we take $z^1=\zeta^1$, $z^2=\zeta^2$, and $z^3=i\theta$, which implies that $\Delta= \bigl(\boldsymbol{\vec{z}\newcdot\vec{z}}\bigr)^{1/2}=\bigl(|\boldsymbol{\vec{\zeta}}|^2-\theta^2\bigr)^{1/2}$.  The resulting formulae reproduce the expressions obtained in \eqs{Lam2d1}{Lam2d2}.

Finally, it is instructive to consider the case of an infinitesimal Lorentz transformation.  Working to linear order in $\boldsymbol{\vec{\zeta}}$ and $\boldsymbol{\vec{\theta}}$, note that
 $\Delta\simeq 0$ in light of \eq{Deltadef}.   Hence, \eqst{LT1}{LT4} reduce to the following result given in block matrix form:
\beqa
\Lambda(\boldsymbol{\vec{\zeta}}\,,\,\boldsymbol{\vec{\theta}}) \simeq 
\left(\begin{array}{c|c}\phm 1 \phm & \,\,\, \zeta^j \\[3pt] \hline \\[-10pt]
\phm\zeta^i\phm & \,\,\,\delta^{ij}-\epsilon^{ijk}\theta^k\end{array}\right),
\eeqa
which coincides with \eq{LamLaminf}.

\section{Reconciling the results of Sections~\ref{sec:fourbyfour} and \ref{sec:twobytwo}}
\label{sec:reconcile}

In this section, we shall verify that the explicit expressions for $\Lambda(\boldsymbol{\vec{\zeta}}\,,\,\boldsymbol{\vec{\theta}})$ obtained, respectively,  in
Sections~\ref{sec:fourbyfour} and \ref{sec:twobytwo} coincide in the general case of non-zero boost and rotation parameters.

First, it is convenient to rewrite \eqs{a}{b} as follows:
\beq \label{a2b2}
a^2=\half\bigl(|\boldsymbol{\vec{\theta}}|^2-|\boldsymbol{\vec{\zeta}}|^2+|\Delta|^2\bigr)\,,\qquad\quad
b^2=\half\bigl(|\boldsymbol{\vec{\zeta}}|^2-|\boldsymbol{\vec{\theta}}|^2+|\Delta|^2\bigr)\,,
\eeq
where $\Delta$ is defined in \eq{Deltadef}.   As noted below \eq{b}, $a$, $b\in\mathbb{R}$ but their undetermined signs have no impact on the expressions obtained for the matrix elements of $\Lambda(\boldsymbol{\vec{\zeta}}\,,\,\boldsymbol{\vec{\theta}})$.   Using \eq{characteristic}, we can fix the relative sign of $a$ and $b$ by choosing
$ab=\boldsymbol{\vec{\theta}}\newcdot\boldsymbol{\vec{\zeta}}$.
It then follows that
\beq \label{bia}
(b+ia)^2=b^2-a^2+2iab=|\boldsymbol{\vec{\zeta}}|^2-|\boldsymbol{\vec{\theta}}|^2+2i\boldsymbol{\vec{\theta}}\newcdot\boldsymbol{\vec{\zeta}}=\Delta^2\,.
\eeq
After taking the positive square root, the signs of $a$ and $b$ are now fixed by identifying
\beq \label{delab}
\Delta=b+ia\,.
\eeq
One can check that \eqst{LT1}{LT4} are unchanged if
$\Delta\to -\Delta$ and/or $\Delta\to\Delta^*$.  This reflects the fact that the expressions obtained for the matrix elements of $\Lambda(\boldsymbol{\vec{\zeta}}\,,\,\boldsymbol{\vec{\theta}})$ do not depend on the choice of signs for $a$ and $b$.

Thus,
\eqst{general}{IAAA} and (\ref{A3}) yield:
\beqa
\begin{array}{lll}
\Lambda^0{}_0&=&\displaystyle\frac{1}{|\Delta|^2}\left[\bigl(b^2-|\boldsymbol{\vec{\zeta}}|^2)\cos a
+\bigl(a^2+|\boldsymbol{\vec{\zeta}}|^2\bigr)\cosh b\right]  \\[13pt]
&=& \half(\cosh b+\cos a)+\displaystyle\frac{|\boldsymbol{\vec{\zeta}}|^2+|\boldsymbol{\vec{\theta}}|^2}{2|\Delta|^2}\bigl(\cosh b-\cos a)\,, \label{cc}
\end{array}
\eeqa
after making use of \eq{a2b2}.
We now employ the following two identities:
\beqa
\cosh b+\cos a &=& 2\cosh\left(\frac{b+ia}{2}\right)\cosh\left(\frac{b-ia}{2}\right)=2\left|\cosh\left(\frac{b+ia}{2}\right)\right|\lsup{2}\,, 
 \label{ident1} \\[6pt]
\cosh b-\cos a &=& 2\sinh\left(\frac{b+ia}{2}\right)\sinh\left(\frac{b-ia}{2}\right)=2\left|\sinh\left(\frac{b+ia}{2}\right)\right|\lsup{2}\,.  
\label{ident2}
\eeqa
Hence, \eqs{delab}{cc} yield
\beq 
\Lambda^0{}_0=|\cosh\bigl(\half\Delta\bigr)|^2+\left|\frac{\sinh\bigl(\half\Delta\bigr)}{\Delta}\right|\lsup{2}\bigl(|\boldsymbol{\vec{\zeta}}|^2+|\boldsymbol{\vec{\theta}}|^2\bigr)\,,
\eeq
in agreement with \eq{LT1}.  

Next, \eqst{general} {IAAA} and (\ref{A3}) yield
\beqa
\begin{array}{lll}
\Lambda^0{}_j&=&\displaystyle\frac{1}{|\Delta|^2}\Biggl\{\left(\displaystyle\frac{b^2}{a}\sin a+\frac{a^2}{b}\sinh b\right)\zeta^j+(\cosh b-\cos a)\epsilon^{jk\ell}\zeta^k\theta^\ell  \\[13pt]
&& \qquad +\left(\displaystyle\frac{\sinh b}{b}-\frac{\sin a}{a}\right)\biggl[\bigl(|\boldsymbol{\vec{\zeta}}|^2-|\boldsymbol{\vec{\theta}}|^2\bigr)\zeta^j+(\boldsymbol{\vec{\theta}\newcdot\vec{\zeta}})\theta^j\biggr]\Biggr\}. \label{aboveeq}
\end{array}
\eeqa
Using \eq{characteristic}, it follows that $|\boldsymbol{\vec{\zeta}}|^2-|\boldsymbol{\vec{\theta}}|^2=b^2-a^2$ and $\boldsymbol{\vec{\theta}\newcdot\vec{\zeta}}=ab$ [the latter with the sign conventions adopted above \eq{bia}].  Inserting these results into \eq{aboveeq}, we obtain
\beq \label{thisresult}
\Lambda^0{}_j=\frac{1}{|\Delta|^2}\biggl[(b\sinh b+a\sin a)\zeta^j+(a\sinh b-b\sin a)\theta^j+(\cosh b-\cos a)\epsilon^{jk\ell}\zeta^k\theta^\ell\biggr]\,.
\eeq

We can rewrite \eq{thisresult} with the help of some identities.  It is straightforward to show that
\beqa
b\sinh b+a\sin a&=& \Delta^*\sinh\bigl(\half\Delta\bigr)\cosh\bigl(\half\Delta^*\bigr)+{\rm c.c.}\,,\label{ba} \\[2pt]
a\sinh b-b\sin a&=&  i\Delta^*\sinh\bigl(\half\Delta\bigr)\cosh\bigl(\half\Delta^*\bigr)+{\rm c.c.}\,, \\[2pt]
(\cosh b-\cos a)\epsilon^{jk\ell}\zeta^k\theta^\ell&=&i \bigl|\sinh\bigl(\half\Delta\bigr)\bigr|\epsilon^{jk\ell}z^{k} z^{*\ell}\,.\label{coshcos}
\eeqa
Collecting the results obtained above, we end up with
\beq
\Lambda^0{}_j= \left(\frac{\sinh\bigl(\half\Delta\bigr)\cosh\bigl(\half\Delta^*\bigr)}{\Delta}\,(\zeta^j+i\theta^j) +{\rm c.c.}\right)+i\,\left|\frac{\sinh\bigl(\half\Delta\bigr)}{\Delta}\right|\lsup{2}\epsilon^{jk\ell}z^{k}z^{*\ell}\,, 
\eeq
in agreement with \eq{LT2}.  

The computation of $\Lambda^i{}_0$ is nearly identical.
The only change is due to the change in the sign multiplying the term proportional to the Levi--Civita tensor. Consequently, it is convenient to replace \eq{coshcos} with an equivalent form:
\beq \label{coshcos2}
(\cosh b-\cos a)\epsilon^{ik\ell}\zeta^k\theta^\ell = -i \bigl|\sinh\bigl(\half\Delta\bigr)\bigr|\epsilon^{ik\ell}z^{*k} z^{\ell}\,.
\eeq
Hence, we end up with
\beqa
\begin{array}{lll}
\Lambda^i{}_0&=&\displaystyle\frac{1}{|\Delta|^2}\biggl[(b\sinh b+a\sin a)\zeta^i+(a\sinh b-b\sin a)\theta^i-(\cosh b-\cos a)\epsilon^{ik\ell}\zeta^k\theta^\ell\biggr]\phantom{xx} \\[13pt]
&=& \left(\displaystyle\frac{\sinh\bigl(\half\Delta\bigr)\cosh\bigl(\half\Delta^*\bigr)}{\Delta}\,(\zeta^i+i\theta^i) +{\rm c.c.}\right)+i\,\left|\displaystyle\frac{\sinh\bigl(\half\Delta\bigr)}{\Delta}\right|\lsup{2}\epsilon^{ik\ell}z^{*k}z^{\ell}\,,
\end{array}
\eeqa 
in agreement with \eq{LT3}. 
\clearpage

Finally, we use \eqst{general}{IAAA} and (\ref{A3}) to obtain
\beqa
\begin{array}{lll}
\Lambda^i{}_j&=&\displaystyle\frac{1}{|\Delta|^2}\Biggl\{(b^2\cos a+a^2\cosh b)\delta^{ij}-\left(\displaystyle\frac{b^2}{a}\sin a+\displaystyle\frac{a^2}{b}\sinh b\right)\epsilon^{ijk}\theta^k  \\[13pt]
&& \qquad\quad +(\cosh b-\cos a)(\zeta^i\zeta^j+\theta^i\theta^j-\delta^{ij}|\boldsymbol{\vec{\theta}}|^2)  \\[8pt]
&& \qquad\quad +\left(\displaystyle\frac{\sinh b}{b}-\displaystyle\frac{\sin a}{a}\right)\biggl[ \epsilon^{ijk}\bigl[(\boldsymbol{\vec{\theta}\newcdot\vec{\zeta}})\zeta^k+\bigl(|\boldsymbol{\vec{\theta}}|^2-|\boldsymbol{\vec{\zeta}}|^2\bigr)\theta^k\bigr]\biggr]\Biggr\}\,.\label{ij}
\end{array}
\eeqa
The following identities can be derived:
\beqa
&&\frac{1}{|\Delta|^2}(\cosh b-\cos a)(\zeta^i\zeta^j+\theta^i\theta^j)=\bigl(z^{*i}z^j+z^i z^{*j}\bigr)\left|\frac{\sinh\bigl(\half\Delta\bigr)}{\Delta}\right|^2\,,\label{zz} \\
&&\frac{1}{|\Delta|^2}\left(\frac{\sinh b}{b}-\frac{\sin a}{a}\right)\boldsymbol{\vec{\theta}\newcdot\vec{\zeta}}=
 i\,\frac{\sinh\bigl(\half\Delta\bigr)}{\Delta}\cosh\bigl(\half\Delta^*\bigr)+{\rm c.c.}\,, \\[5pt]
&&\frac{1}{|\Delta|^2}\left[b^2\cos a+a^2\cosh b-(\cosh b-\cos a)|\boldsymbol{\vec{\theta}}|^2\right] 
=|\cosh\bigl(\half\Delta\bigr)|^2-\left|\frac{\sinh\bigl(\half\Delta\bigr)}{\Delta}\right|^2\!\!\bigl(|\boldsymbol{\vec{\zeta}}|^2+|\boldsymbol{\vec{\theta}}|^2\bigr), \label{dij}  \nonumber \\[-10pt]
&& \phantom{line} \\
&&\frac{1}{|\Delta|^2}\Biggl\{\left(\frac{\sinh b}{b}-\frac{\sin a}{a}\right)\bigl(|\boldsymbol{\vec{\theta}}|^2
-|\boldsymbol{\vec{\zeta}}|^2\bigr)-\left(\frac{b^2}{a}\sin a+\frac{a^2}{b}\sinh b\right)\Biggr\}  \nonumber \\
&& \qquad\qquad\qquad\qquad\qquad\qquad =-\left\{\frac{\sinh\bigl(\half\Delta\bigr)}{\Delta}\cosh\bigl(\half\Delta^*\bigr)+{\rm c.c.}\right\}. 
\eeqa
Note that the terms proportional to $\epsilon^{ijk}$ in \eq{ij} combine nicely and yield
\beq \label{nice}
\frac{i\sinh\bigl(\half\Delta\bigr)\cosh\bigl(\half\Delta^*\bigr)}{\Delta}\,\epsilon^{ijk}z^k+{\rm c.c.}\,,
\eeq
after putting $z^k=\zeta^k+i\theta^k$.

Collecting the results obtained above, we end up with
\beqa
\begin{array}{lll}
\Lambda^i{}_j&=&\biggr\{|\cosh\bigl(\half\Delta\bigr)|^2-\left|\displaystyle\frac{\sinh\bigl(\half\Delta\bigr)}{\Delta}\right|^2\bigl(|\boldsymbol{\vec{\zeta}}|^2+|\boldsymbol{\vec{\theta}}|^2\bigr)\biggl\}\delta^{ij}+(z^{*i}z^j+z^i z^{*j})\left|\displaystyle\frac{\sinh\bigl(\half\Delta\bigr)}{\Delta}\right|\lsup{2}
  \\[16pt]
&& \qquad\qquad
+\left(\displaystyle\frac{i\sinh\bigl(\half\Delta\bigr)\cosh\bigl(\half\Delta^*\bigr)}{\Delta}\epsilon^{ijk}z^k+{\rm c.c.}\right)\,,
\end{array}
\eeqa
in agreement with \eq{LT4}.

We have therefore verified by an explicit computation that the results obtained in \linebreak\mbox{\eqst{general}{f23}} are equivalent to \eqst{LT1}{LT4}.  
In particular, we have established that
\beq \label{theorem}
\Lambda^\mu{}_\nu(\boldsymbol{\vec{\zeta}},\boldsymbol{\vec{\theta}})=\half\Tr\bigl(M^\dagger\sigmabar^\mu M\sigma_\nu\bigr)\,,
\eeq
where $M=\exp\left\{-\half(\boldsymbol{\vec{\zeta}}+i\boldsymbol{\vec{\theta}})\boldsymbol{\newcdot\vec{\sigma}}\right\}$.

\section{Final remarks}
\label{final}

The main goal of this paper is to exhibit an explicit form for the $4\times 4$ proper orthochronous Lorentz transformation matrix as a function of general boost and rotation parameters
$\boldsymbol{\vec{\zeta}}$ and~$\boldsymbol{\vec{\theta}}$.   Whereas the matrices $\Lambda(\boldsymbol{\vec{\zeta}},\boldsymbol{\vec{0}})$ and
$\Lambda(\boldsymbol{\vec{0}},\boldsymbol{\vec{\theta}})$ are well known and appear in many textbooks, the explicit form for more general $\Lambda(\boldsymbol{\vec{\zeta}},\boldsymbol{\vec{\theta}})$ is much less well known.  Two different derivations are provided for $\Lambda(\boldsymbol{\vec{\zeta}},\boldsymbol{\vec{\theta}})$.  One derivation evaluates the exponential of a real $4\times 4$ matrix $A$ that satisfies $(GA)^{\T}=-GA$ [where $G\equiv{\rm diag}(1,-1,-1,-1)$], and a second derivation evaluates $\half\Tr\bigl(M^\dagger\sigmabar^\mu M\sigma_\nu\bigr)$, where the $2\times 2$ matrix $M=\exp\{-\half(\boldsymbol{\vec{\zeta}}+i\boldsymbol{\vec{\theta}})\boldsymbol{\newcdot\vec{\sigma}}\}$.  Although the results obtained by the two computations look somewhat different at first, we have verified by explicit calculation that these two results are actually equivalent.

One can also obtain the most general proper orthochronous Lorentz transformation in another way by invoking
the following theorem (e.g., see Section 1.5 of Ref.~\cite{Sexl}, Section 6.6 of Ref.~\cite{Rao}, or Section 4.5 of Ref.~\cite{Scheck}):

\begin{changemargin}{3em}{3em}
Every proper orthochronous Lorentz transformation $\Lambda(\boldsymbol{\vec{\zeta}},\boldsymbol{\vec{\theta}})$ possesses a unique factorization into a product of a boost and a rotation in two different ways:
\beq \label{polar}
\Lambda(\boldsymbol{\vec{\zeta}},\boldsymbol{\vec{\theta}})=\Lambda(\boldsymbol{\vec{\zeta}^\prime},\boldsymbol{\vec{0}})
\Lambda(\boldsymbol{\vec{0}},\boldsymbol{\vec{\theta}^\prime})=\Lambda(\boldsymbol{\vec{0}},\boldsymbol{\vec{\theta}^{\prime\prime}})\Lambda(\boldsymbol{\vec{\zeta}^{\prime\prime}},\boldsymbol{\vec{0}}).
\eeq
for an appropriate choice of parameters $\{\boldsymbol{\vec{\zeta}^\prime}$, $\boldsymbol{\vec{\theta}^\prime}\}$ and $\{\boldsymbol{\vec{\zeta}^{\prime\prime}},  \boldsymbol{\vec{\theta}^{\prime\prime}}\}$, respectively.\footnote{\Eq{polar} is called the polar decomposition of SO$_0$(1,3) in Refs.~\cite{Gallier,Moretti,Urbantke}.}
\end{changemargin}

\noindent
In particular, if none of the parameters are zero, then $\boldsymbol{\vec{\zeta}}\neq\boldsymbol{\vec{\zeta}^\prime}\neq \boldsymbol{\vec{\zeta}^{\prime\prime}}$ and
$\boldsymbol{\vec{\theta}}\neq\boldsymbol{\vec{\theta}^\prime}\neq \boldsymbol{\vec{\theta}^{\prime\prime}}$ due to the fact that boosts and rotations do not commute
[as a consequence of the commutation relations given in \eq{sskk}].  Indeed, for non-vanishing boost and rotation parameters,
\beq \label{noncommute}
\Lambda(\boldsymbol{\vec{\zeta}}\,,\,\boldsymbol{\vec{\theta}})=  \exp\bigl(
-i\mathbold{{\vec\theta}\newcdot}\boldsymbol{\vec s}-i\mathbold{{\vec\zeta}\newcdot}\boldsymbol{\vec k}\bigr)\neq \exp\bigl(
-i\mathbold{{\vec\theta}\newcdot}\boldsymbol{\vec s}\bigr)\exp\bigl(-i\mathbold{{\vec\zeta}\newcdot}\boldsymbol{\vec k}\bigr)
\neq\exp\bigl(-i\mathbold{{\vec\zeta}\newcdot}\boldsymbol{\vec k}\bigr)\exp\bigl(
-i\mathbold{{\vec\theta}\newcdot}\boldsymbol{\vec s}\bigr)\,.
\eeq
In contrast to \eq{noncommute}, when considering infinitesimal Lorentz transformations,
the boost matrix [\eq{Lmatrixzetainf}] and the rotation matrix [\eq{Lmatrix2inf}] commute at linear order, which results in \eq{LamLaminf}.  The effects of the noncommutativity appear first at quadratic order in the boost and rotation parameters.

Given the parameters $\{\boldsymbol{\vec{\zeta}^\prime}$, $\boldsymbol{\vec{\theta}^\prime}\}$ (or $\{\boldsymbol{\vec{\zeta}^{\prime\prime}}$,  $\boldsymbol{\vec{\theta}^{\prime\prime}}\}$), it would be quite useful to be able to obtain expressions for the corresponding parameters of $\Lambda(\boldsymbol{\vec{\zeta}}, \boldsymbol{\vec{\theta}})$.
The formulae that determine  $\{\boldsymbol{\vec{\zeta}}, \boldsymbol{\vec{\theta}}\}$ in \eq{polar} are quite complicated~\cite{KKZ}, although they could in principle be derived by using the explicit matrix representations given in this paper.  This is left as an exercise for the reader.

\section*{Acknowledgments}

I am grateful to Jo\~ao P.~Silva for discussions in which he challenged me to provide an explicit proof of \eq{theorem} and for his encouragements during the writeup of this work.
HEH is partially supported in part by the U.S. Department of Energy Grant number~\uppercase{DE-SC}0010107.

\begin{appendices}

\section{\texorpdfstring{Rodrigues' rotation formula}{Appendix A: Rodrigues' rotation formula}}
\label{appA}
\renewcommand{\theequation}{A.\arabic{equation}}
\setcounter{equation}{0}

A proper rotation matrix 
$R(\axis,\theta)$ [which satisfies $RR^{\T}=\id_3$ and $\det R=1$] represents an \textit{active} transformation consisting of a counterclockwise rotation by an angle $\theta$ about an axis~$\axis$
with respect to a fixed Cartesian coordinate system.
For example, the matrix representation of the counterclockwise rotation by an angle $\theta$ about
the $z$-axis is given by
\beq \label{z}
R(\boldsymbol{\hat{z}},\theta)\equiv\pmat{\cos\theta & \,\,\, -\sin\theta & \,\,\, \phm 0\\ \sin\theta & \,\,\ \phm\!\cos\theta & \,\,\,\phm  0
\\ 0 & \,\,\,\phm 0 & \,\,\,\phm 1}\,.
\eeq

The matrix elements of $R(\axis,\theta)$ will be denoted by $R_{ij}$, where the indices of the tensors in this Appendix are written in the lowered position to
simplify the typography of the presentation.
The goal of this Appendix is to provide a simple derivation of Rodrigues' formula for an active (counterclockwise) rotation by an angle $\theta$ about an axis
that points along the unit vector $\boldsymbol{\hat{n}}=(n_1\,,\,n_2\,,\,n_3)$.  Note that since 
$\axis$
is a unit vector, it follows that
\beq \label{n}
n_1^2+n_2^2+n_3^2=1\,.
\eeq

The traditional approach to deriving Rodrigues' rotation formula involves the computation of 
the exponential of an arbitrary $3\times 3$ real antisymmetric matrix (e.g., see Refs.~\cite{Marsden,Gallier}).
Below, we provide an alternative derivation of the formula for $R_{ij}$ that makes use of the techniques of
tensor algebra.  

Consider how $R_{ij}$ changes under an orthogonal change of basis,
which can be viewed as a orthogonal transformation of the coordinate axes.  Using the well-known results derived in any textbook
on matrices and linear algebra, one can check that the transformation of the components of $R_{ij}$ under a change of 
basis corresponds to the transformation law of
a second-rank Cartesian tensor.  
Likewise, the~$n_i$ are components of a vector (equivalently, a first-rank tensor).
Two other important quantities of the analysis are the \textit{invariant} tensors $\delta_{ij}$ (the Kronecker delta)
and $\epsilon_{ijk}$ (the Levi--Civita tensor).
 If we invoke the covariance of Cartesian tensor equations, then
one must be able to express $R_{ij}$ in terms of a second-rank tensor composed of $n_i$, $\delta_{ij}$ and
$\epsilon_{ijk}$, as there are no other tensors in the problem that could provide a source of indices.
Thus, the form of the formula for $R_{ij}$ must be
\beq \label{rij}
R_{ij}=a\,\delta_{ij}+b\,n_i n_j+c\,\epsilon_{ijk}n_k\,,
\eeq
where there is an implicit sum over the repeated index $k$ in the last term of \eq{rij}.
The numbers $a$, $b$ and $c$ are real scalar quantities.
As such, $a$, $b$ and $c$ are functions of $\theta$, since the rotation
angle is the only scalar variable
in this problem.

We now determine the conditions that are satisfied by $a$, $b$ and
$c$.  The first condition is obtained by noting that
\beq \label{eigenvector}
R(\axis,\theta)\axis=\axis\,.
\eeq
This is clearly true, since $R(\axis,\theta)$, when acting on a vector, rotates the vector around the axis~$\axis$, whereas
any vector parallel to the axis of rotation
is invariant under the action of $R(\axis,\theta)$.  In terms of components,
\beq \label{rn}
R_{ij}n_j=n_i\,.
\eeq
To determine the consequence of this equation, we insert \eq{rij} into \eq{rn}.
In light of \eq{n}, it follows immediately that $n_i(a+b)=n_i$.  Hence,
\beq \label{ab}
a+b=1\,.
\eeq

Since the formula for $R_{ij}$ given by \eq{rij} must be completely general, it must hold
for any special case.  In particular, consider the case where $\axis=\boldsymbol{\hat{z}}$.  In this case, \eqs{z}{rij} yield
\beq \label{rac}
R(\boldsymbol{k},\theta)_{11}=\cos\theta=a\,,\qquad\qquad
R(\boldsymbol{k},\theta)_{12}=-\sin\theta=c\,,
\eeq
after using $n_3=\epsilon_{123}=1$.
Consequently, \eqs{ab}{rac} yield
\beq\label{ct}
a=\cos\theta\,,\qquad\quad b= 1-\cos\theta\,,\qquad\quad c=-\sin\theta\,.
\eeq
Inserting these results into \eq{rij}, we obtain 
Rodrigues' rotation formula:
\beq \label{Rij}
R_{ij}(\axis,\theta)
=\cos\theta\,\delta_{ij}+(1-\cos\theta)n_i n_j-\sin\theta\,\epsilon_{ijk}n_k\,.
\eeq

Note that
\beqa
R(\axis,\theta+2\pi k)&=&R(\axis,\theta)\,,\qquad k=0,\pm 1\,\pm 2\,\ldots\,,\label{rnk} \\[8pt]
[R(\axis,\theta)]^{-1}&=&R(\axis,-\theta)=R(-\axis,\theta)\,.\label{Rrel}
\eeqa
Combining these two results, it follows that
\beq \label{rnt}
R(\axis,2\pi-\theta)=R(-\axis,\theta)\,,
\eeq
which implies that any three-dimensional proper rotation can be described by a counterclockwise rotation 
by an angle $\theta$ about some axis $\axis$,
where $0\leq\theta\leq\pi$.

\section{\texorpdfstring{$\eta^\dagger\sigmabar^\mu\chi$}{Appendix B: \texteta\textCross\textsigma\overbar\superu\textchi} transforms as a Lorentz four-vector} 
\label{appB}
\renewcommand{\theequation}{B.\arabic{equation}}
\setcounter{equation}{0}

\Eq{assert} asserts that the spinor product $\eta^\dagger\sigmabar^\mu\chi$ transforms as a Lorentz four-vector.   In light of \eq{under}, it follows that \eq{Lam} must be satisfied (and vice versa).   In this Appendix, we shall establish \eq{Lam} by demonstrating that both sides of this identity agree to first order in $\boldsymbol{\vec{\zeta}}$ and~$\boldsymbol{\vec{\theta}}$.  

In addition to the $\sigma^{\mu\nu}$ defined in \eq{sigmamunu}, it is convenient to introduce the set of $2\times 2$ matrices, 
\beq \label{sigmabarmunu}
\sigmabar^{\mu\nu}=-\sigmabar^{\nu\mu}\equiv \tfrac14 i(\sigmabar^\mu\sigma^\nu-\sigmabar^\nu\sigma^\mu)\,.
\eeq
Then, using the properties of the Pauli matrices, \eqs{emdef}{M} yield 
\beq \label{Md}
M^\dagger=\exp\left(\half i\theta_{\rho\lambda}\sigmabar^{\rho\lambda}\right)=\exp\left(\half i\boldsymbol{\vec{\theta}\newcdot\vec{\sigma}}-\half\boldsymbol{\vec{\zeta}\newcdot\vec{\sigma}}\right)\,.
\eeq

Working to first order in the parameters $\theta_{\rho\lambda}$ and making use of eqs.~(\ref{explicitsmunu}), (\ref{lambda44again}), (\ref{M}), and (\ref{Md}),
\beqa
\Lambda^\mu{}_\nu &\simeq&\delta^\mu_\nu+\half\left(\theta_{\lambda\nu}
\eta^{\lambda\mu}-\theta_{\nu\rho}\eta^{\rho\mu} \right)\,,\label{Linf}\\
M& \simeq& \boldsymbol{I}_2
-\half i\theta_{\rho\lambda}\sigma^{\rho\lambda}\,, \label{Minf} \\[4pt]
M^\dagger &\simeq &\boldsymbol{I}_2+\half i
\theta_{\rho\lambda}\sigmabar^{\rho\lambda}\,.\label{MDinf}
\eeqa
It then follows that
\beq \label{MsigM}
M^\dagger\sigmabar^\mu M\simeq \bigl(\boldsymbol{I}_2+\half i
\theta_{\rho\lambda}\sigmabar^{\rho\lambda}\bigr)\sigmabar^\mu\bigl(\boldsymbol{I}_2
-\half i\theta_{\rho\lambda}\sigma^{\rho\lambda}\bigr)\simeq \sigmabar^\mu+\half i\theta_{\rho\lambda}\bigl(\sigmabar^{\rho\lambda}\sigmabar^\mu-\sigmabar^\mu\sigma^{\rho\lambda}\bigr)\,.
\eeq
One can easily derive the following identity~\cite{Dreiner:2008tw,Dreiner:2023yus}:
\beq 
\sigmabar^{\rho\lambda}\sigmabar^\mu-\sigmabar^\mu\sigma^{\rho\lambda}=i\bigl(\eta^{\lambda\mu}\sigmabar^\rho-\eta^{\rho\mu}\sigmabar^\lambda\bigr)\,.
\eeq
Hence, \eq{MsigM} yields
\beqa
\begin{array}{lll}
M^\dagger\sigmabar^\mu M & \simeq & \sigmabar^\mu-\half \theta_{\rho\lambda}\bigl(\eta^{\lambda\mu}\sigmabar^\rho-\eta^{\rho\mu}\sigmabar^\lambda\bigr) 
\simeq  \bigl[\delta^\mu_\nu-\half\theta_{\rho\lambda}\bigl(\eta^{\lambda\mu}\delta^\rho_\nu-\eta^{\rho\mu}\delta^\lambda_\nu\bigr)\bigr]\sigmabar^\nu  \\[10pt]
&\simeq & \bigl[\delta^\mu_\nu-\half\bigl(\theta_{\nu\lambda} \eta^{\lambda\mu}-\theta_{\rho\nu}\eta^{\rho\mu}\bigr)\bigr]\sigmabar^\nu \simeq 
\bigl[\delta^\mu_\nu+\half\bigl(\theta_{\lambda\nu} \eta^{\lambda\mu}-\theta_{\nu\rho}\eta^{\rho\mu}\bigr)\bigr]\sigmabar^\nu=\Lambda^\mu{}_\nu\sigmabar^\nu\,,\phantom{xxx}\label{MdsigM}
\end{array}
\eeqa
after using the antisymmetry of $\theta_{\nu\lambda}$ in the penultimate step above.  
After employing \eq{Linf}
in the final step above, we conclude that
\beq \label{MMLam}
M^\dagger\sigmabar^\mu M =\Lambda^\mu{}_\nu\sigmabar^\nu\,,
\eeq
thereby confirming \eq{Lam}.  
In particular, it follows that $\eta^\dagger\sigmabar^\mu\chi$ transforms as a Lorentz four-vector
in light of \eqs{under}{assert}, as previously noted.
\Eq{MMLam} is a statement of the well-known isomorphism SO(1, 3)$_0\iso$ SL(2, $\mathbb{C}$)/$\mathbb{Z}_2$, since the SL(2, $\mathbb{C})$ matrices $M$ and $-M$ correspond to the same Lorentz transformation $\Lambda$.

Of course, the derivation of \eq{MMLam} is much simpler than a direct derivation of \eq{L4}, which requires the explicit evaluation of all the relevant matrix exponentials.   
Indeed, we can assert that having derived \eq{MMLam} to first order in $\theta_{\rho\lambda}$, this result must be true for arbitrary $\theta_{\rho\lambda}$.
The reason that a derivation based on the infinitesimal forms of $\Lambda$, $M$ and $M^\dagger$
is sufficient is due to the strong constraints imposed by the group multiplication law of the Lorentz group near the identity element, which in light of the discussion following \eq{exponentiate} implies that a proper orthochronous Lorentz transformation can be expressed as an exponential of an element of the corresponding Lie algebra.
 
There is a second inequivalent two-dimensional matrix representation
of ${\rm SL}(2,\mathbb{C})$ whose general element is represented by the matrix $(M^{-1})^\dagger$, as discussed in greater detail in Refs.~\cite{Dreiner:2008tw,Dreiner:2023yus}.
This leads to a second identity that is similar to that of  \eq{MMLam}:
\beq \label{secondid}
M^{-1}\sigma^\mu (M^{-1})^\dagger=\Lambda^\mu{}_\nu\,\sigma^\nu\,.
\eeq
One can derive \eq{secondid} by again working to first order in the parameters $\theta_{\rho\lambda}$ and making use of \eqst{Linf}{MDinf}:
\beq
M^{-1}\sigma^\mu (M^{-1})^\dagger \!\simeq\!  \bigl(\boldsymbol{I}_2+\half i
\theta_{\rho\lambda}\sigma^{\rho\lambda}\bigr)\sigma^\mu\bigl(\boldsymbol{I}_2
-\half i\theta_{\rho\lambda}\sigmabar^{\rho\lambda}\bigr) \simeq \sigma^\mu+\half i\theta_{\rho\lambda}\bigl(\sigma^{\rho\lambda}\sigma^\mu-\sigma^\mu\sigmabar^{\rho\lambda}\bigr).
\eeq
In light of the identity~\cite{Dreiner:2008tw,Dreiner:2023yus},
\beq 
\sigma^{\rho\lambda}\sigma^\mu-\sigma^\mu\sigmabar^{\rho\lambda}=i\bigl(\eta^{\lambda\mu}\sigma^\rho-\eta^{\rho\mu}\sigma^\lambda\bigr)\,,
\eeq
it follows that
\beqa
\begin{array}{lll}
&&\hspace{-0.45in}  M^{-1}\sigma^\mu (M^{-1})^\dagger  \simeq   \sigma^\mu-\half \theta_{\rho\lambda}\bigl(\eta^{\lambda\mu}\sigma^\rho-\eta^{\rho\mu}\sigma^\lambda\bigr) 
\simeq  \bigl[\delta^\mu_\nu-\half\theta_{\rho\lambda}\bigl(\eta^{\lambda\mu}\delta^\rho_\nu-\eta^{\rho\mu}\delta^\lambda_\nu\bigr)\bigr]\sigma^\nu  \\[10pt]
&& \quad\!\!\! \simeq  \bigl[\delta^\mu_\nu-\half\bigl(\theta_{\nu\lambda} \eta^{\lambda\mu}-\theta_{\rho\nu}\eta^{\rho\mu}\bigr)\bigr]\sigma^\nu \simeq 
\bigl[\delta^\mu_\nu+\half\bigl(\theta_{\lambda\nu} \eta^{\lambda\mu}-\theta_{\nu\rho}\eta^{\rho\mu}\bigr)\bigr]\sigma^\nu=\Lambda^\mu{}_\nu\,\sigma^\nu,
\end{array}
\eeqa
which establishes \eq{secondid} after employing \eq{Linf} in the final step above.
\clearpage

Multiplying \eq{secondid} on the right by $\sigmabar_\rho$ and using $\Tr(\sigma^\nu\sigmabar_\rho)=2\delta^\nu_{\rho}$, it \mbox{follows that}
\beq \label{Lamagain}
\Lambda^\mu{}_\nu=\half\Tr\bigl[M^{-1}\sigma^\mu (M^{-1})^\dagger\sigmabar_\nu\bigr]\,,
\eeq
which provides yet another formula for the most general orthochronous Lorentz transformation matrix.
Using block matrix notation, \eq{Lamagain} yields
\beq  \label{Lmatrix4}
\Lambda(\boldsymbol{\vec{\zeta}}\,,\,\boldsymbol{\vec{\theta}}) =
\left(\begin{array}{c|c}\Lambda^0{}_0  & \, \Lambda^0{}_j\\[3pt] \hline \\[-10pt]
\Lambda^i{}_0 & \,\Lambda^i{}_j\end{array}\right)=
\frac12\left(\begin{array}{c|c}\Tr\bigl[M^{-1}(M^{-1})^\dagger\bigr]\ & \,\,\, \Tr \bigl[(M^{-1})^\dagger\sigma^j M^{-1}\bigr] \\[3pt] \hline \\[-10pt]
\Tr \bigl[M^{-1}\sigma^i (M^{-1})^\dagger\bigr]\ & \,\,\, \Tr\bigl[M^{-1} \sigma^i (M^{-1})^\dagger\sigma^j\bigr] \end{array}\right),
\eeq
after noting that $\sigmabar_j=-\sigma_j=\sigma^j$ [cf.~\eq{sigmadefs}].
Comparing with \eq{Lmatrix3}, we see that \eq{Lmatrix4} is obtained by changing $M\to (M^{-1})^\dagger$ and
$M^\dagger\to M^{-1}$, which results in $\boldsymbol{\vec{\theta}}\to \boldsymbol{\vec{\theta}}$ and
$\boldsymbol{\vec{\zeta}}\to -\boldsymbol{\vec{\zeta}}$, or equivalently $\boldsymbol{\vec{z}}\to -\boldsymbol{\vec{z}}\lsup{\,*}$ and  $\Delta\to \Delta^*$.  
In addition, the block off-diagonal elements of $\Lambda(\boldsymbol{\vec{\zeta}}\,,\,\boldsymbol{\vec{\theta}})$ have changed sign.  Under these replacements, it is straightforward to check that the resulting expressions for $\Lambda^\mu{}_\nu$ are the same as those 
obtained previously in \eqst{LT1}{LT4}.   That is, \eq{Lamagain} is established by explicit calculation.

\section{\texorpdfstring{$\overline{\Psi}\gamma^\mu\Psi$}{Appendix C: \uppercasepsi\overbar\textgamma\superu\uppercasepsi} transforms as a Lorentz four-vector}
\label{appC}
\renewcommand{\theequation}{C.\arabic{equation}}
\setcounter{equation}{0}

Most textbook treatments of the Dirac equation employ the more familiar four-component spinors and Dirac gamma matrices (e.g., see Ref.~\cite{Peskin}).   
The relation between the two-component and four-component spinor formalisms is briefly presented in this Appendix.  Further details can be found in Refs.~\cite{Dreiner:2008tw,Dreiner:2023yus}.  

One can construct four-component spinors
\beq
\Psi\equiv\begin{pmatrix}\chi\\[4pt]
\eta^\dagger \end{pmatrix}\,,
\label{general4comp}
\eeq
in terms of a pair of two-component spinors $\chi$ and $\eta$.  The Dirac gamma matrices are defined via their anticommutation relations:
\beq 
\label{antic}
\{\gamma^\mu\,,\,\gamma^\nu\}\equiv \gamma^\mu\gamma^\nu+\gamma^\nu\gamma^\mu=2\eta^{\mu\nu}\,.
\eeq

In the so-called chiral representation of the gamma matrices, 
\beq \label{gamma4}
\gamma^\mu = \begin{pmatrix} 0&\,\,\, \sigma^\mu
\\ \sigmabar^{\mu} &\,\,\,  0\end{pmatrix}\,.
\eeq
It is convenient to introduce
\beq \label{Sigmamunu}
\half\Sigma^{\mu\nu}\equiv\tfrac{1}{4} i [\gamma^\mu,\gamma^\nu]=
\begin{pmatrix} \sigma^{\mu\nu}& \,\,\, 0\\
0 & \,\,\,  \sigmabar^{\mu\nu}\end{pmatrix}\,,
\eeq
where $[\gamma^\mu,\gamma^\nu]\equiv \gamma^\mu\gamma^\nu-\gamma^\nu\gamma^\mu$.
The Dirac adjoint spinor is defined by
\beq \label{adjoint}
\overline{\Psi}(x)\equiv\Psi^\dagger(x) \gamma^0=
\begin{pmatrix}\eta & \chi^\dagger
\end{pmatrix}\,.
\eeq
The matrices $\gamma^\mu$ and $\Sigma^{\mu\nu}$ satisfy
\beqa
\gamma^0 \gamma^\mu \gamma^0&=&(\gamma^\mu)^\dagger\,,\label{gammazero}\\[5pt]
\gamma^0 \Sigma^{\mu\nu} \gamma^0&=&(\Sigma^{\mu\nu})^\dagger\,.\label{sigmunu}
\eeqa

Four-component spinors transform under an active Lorentz transformation as
\beq \label{PPbar}
\Psi^\prime=\mathds{M}\Psi\,,
\eeq
where
\beq \label{dsM}
\mathds{M}\equiv\begin{pmatrix} M & \quad 0 \\
0 & \quad (M^{-1})^\dagger \end{pmatrix}
=\exp\left(-\quarter i\theta_{\mu\nu}\Sigma^{\mu\nu}\right)
\eeq
combines the two inequivalent two-dimensional matrix representations of ${\rm SL}(2,\mathbb{C})$, 
\beqa
M&=&\exp\left(-\half i\theta_{\rho\lambda}\sigma^{\rho\lambda}\right)=\exp\left(-\half i\boldsymbol{\vec{\theta}\newcdot\vec{\sigma}}-\half\boldsymbol{\vec{\zeta}\newcdot\vec{\sigma}}\right)\,, \\[6pt]
(M^{-1})^\dagger&=&\exp\left(-\half i\theta_{\rho\lambda}\sigmabar^{\rho\lambda}\right)=\exp\left(-\half i\boldsymbol{\vec{\theta}\newcdot\vec{\sigma}}+\half\boldsymbol{\vec{\zeta}\newcdot\vec{\sigma}}\right)\,.
\eeqa
To compute the corresponding matrix inverses, simply change the overall sign of the parameters~$\theta_{\mu\nu}$.  For example,
\beq
\mathds{M}^{-1}=\exp\left(\quarter i\theta_{\mu\nu}\Sigma^{\mu\nu}\right)\,.
\eeq

In light of \eq{sigmunu}, one can easily check that the $4\times 4$ matrix $\mathds{M}$ satisfies
\beq \label{gamMgam}
\gamma^0 \mathds{M}\gamma^0=(\mathds{M}^{-1})^\dagger\,.
\eeq
Using \eqs{adjoint}{PPbar}, it then follows that
\beq
\overline\Psi\lsup{\,\prime}\equiv\Psi^{\prime \dagger}\gamma^0=\Psi^\dagger\mathds{M}^\dagger\gamma^0=\overline{\Psi}\gamma^0\mathds{M}^\dagger\gamma^0\,.
\eeq
Finally, taking the hermitian conjugate of \eq{gamMgam} and using \eq{gammazero} [which implies that $(\gamma^0)^\dagger=\gamma^0$ in light of \eq{antic}], we end up with
\beq \label{PPbar2}
\overline\Psi\lsup{\,\prime}=\overline{\Psi}\mathds{M}^{-1}\,,
\eeq
under an active Lorentz transformation.

It is now straightforward to verify that the 
identities, \eqs{MMLam}{secondid}, derived in Appendix~\ref{appB}, 
are equivalent to
 \beq \label{sgammas}
 \mathds{M}^{-1}\gamma^\mu  \mathds{M}=\Lambda^\mu{}_\nu\gamma^\nu\,,
\eeq
after employing \eqs{gamma4}{dsM}.  Consequently, in light of \eqss{PPbar}{PPbar2}{sgammas}, it follows that under an active Lorentz transformation,
\beq
\overline\Psi\gamma^\mu\Psi\longrightarrow \overline\Psi\mathds{M}^{-1}\gamma^\mu\mathds{M}\Psi=\Lambda^\mu{}_\nu\overline\Psi\gamma^\nu\Psi\,.
\eeq
That is, under a Lorentz transformation, $\overline\Psi\gamma^\mu\Psi$ transforms as a four-vector.  Moreover, using
$\Tr(\gamma^\mu\gamma_\nu)=4\delta^\mu_\nu$, \eq{sgammas} yields
\beq \label{another}
\Lambda^\mu{}_\nu=\tfrac14 \Tr(\mathds{M}^{-1}\gamma^\mu\mathds{M}\gamma_\nu)\,.
\eeq
Of course, \eq{another} is equivalent to \eqs{L4}{Lamagain} taken together.  

Note that using \eq{another} to obtain an explicit form for $\Lambda^\mu{}_\nu$ requires the evaluation of the trace of a product of four $4\times 4$ matrices.   In contrast, the computation of $\Lambda^\mu{}_\nu$ presented in Section~\ref{sec:twobytwo} is more straightforward involving less duplication of effort.

\end{appendices}

\end{document}